\documentclass[conference]{IEEEtran}
\IEEEoverridecommandlockouts
\usepackage{url}

\usepackage{cite}
\usepackage{amsmath,amssymb,amsfonts}
\usepackage{algorithmic}
\usepackage{graphicx}
\usepackage{textcomp}
\usepackage{xcolor}
\usepackage[ruled,vlined]{algorithm2e} % Add this line to include the algorithm2e package
\usepackage{enumitem}

\usepackage{lipsum}  % 示例文本
\usepackage{fancyhdr}

\usepackage{float}
\usepackage{diagbox}
\usepackage{comment}
\def\BibTeX{{\rm B\kern-.05em{\sc i\kern-.025em b}\kern-.08em
    T\kern-.1667em\lower.7ex\hbox{E}\kern-.125emX}}
\begin{document}

\title{%DEthna: Accurate Ethereum Network Topology  Discovery with Marked Transactions%\\
ExClique: An Express Consensus Algorithm for High-Speed Transaction Process in Blockchains
%\thanks{The corresponding author is Shengli Zhang (zsl@szu.edu.cn).}
\thanks{The corresponding authors are Shengli Zhang and Shiting Wen. The research is supported in part by National Natural Science Foundation of China (62171291), in part by Shenzhen Key Research Project (JSGG20220831095603007), in part by Zhejiang Provincial Natural Science Foundation of China (No. LY24F020021), in part by Ningbo Science and Technology Special Projects (No. 2022Z095, No. 2022Z235), and in part by Australian Research Council (ARC) Discovery Project (DP210101723).}
%\thanks{The corresponding author is Shengli Zhang. The research is partially supported by National Natural Science Foundation of China (62171291), Shenzhen Key Research Project (JSGG20220831095603007, JCYJ20220818100810023, JCYJ20220818101609021) and Science and Technology Program (JCYJ20210324094609027), Australian Research Council (ARC) Discovery Project (DP210101723), and Hong Kong RGC Research Impact Fund (No. R5060-19, No. R5034-18), Areas of Excellence Scheme (AoE/E-601/22-R) and  General Research Fund (No. 152203/20E, 152244/21E, 152169/22E, 152228/23E).}
%\thanks{The corresponding author is Shengli Zhang (zsl@szu.edu.cn). The research is supported in part by the National Natural Science Foundation of China (62171291), in part by the Shenzhen Key Research Project (JSGG20220831095603007, JCYJ20220818100810023, JCYJ20220818101609021), in part by the Shenzhen Science and Technology Program (JCYJ20210324094609027), in part by the Australian Research Council (ARC) Discovery Project (DP210101723), and in part by the Hong Kong RGC Research Impact Fund (No. R5060-19, No. R5034-18), Areas of Excellence Scheme (AoE/E-601/22-R) and  General Research Fund (No. 152203/20E, 152244/21E, 152169/22E, 152228/23E).}
}

\DeclareRobustCommand*{\IEEEauthorrefmark}[1]{%
    \raisebox{0pt}[0pt][0pt]{\textsuperscript{\footnotesize\ensuremath{#1}}}}
\author{\IEEEauthorblockN{Chonghe Zhao\IEEEauthorrefmark{1}, Yipeng Zhou\IEEEauthorrefmark{2}, Shengli Zhang\IEEEauthorrefmark{1},  Quan Z. Sheng\IEEEauthorrefmark{2}, Yang Zhang\IEEEauthorrefmark{3}, Shiting Wen\IEEEauthorrefmark{4}}
\IEEEauthorblockA{\IEEEauthorrefmark{1}College of
Electronic and Information Engineering, Shenzhen University, Shenzhen, China\\
\IEEEauthorrefmark{2}School of Computing, Macquarie University, Sydney, Australia\\
\IEEEauthorrefmark{3}Anuradha and Vikas Sinha Department of Data Science, University of North Texas, Denton, Texas, USA \\
\IEEEauthorrefmark{4}School of Computer and Data Engineering, NingboTech University, Ningbo, China\\
Email: zhaochonghe\_szu@163.com, yipeng.zhou@mq.edu.au, zsl@szu.edu.cn\\ michael.sheng@mq.edu.au, yang.zhang@unt.edu, wensht@nbt.edu.cn}
}

% \and
% \and

% \author{\IEEEauthorblockN{Chonghe Zhao\IEEEauthorrefmark{2}, Yipeng Zhou\IEEEauthorrefmark{3}, Shengli Zhang\IEEEauthorrefmark{2}, Taotao Wang\IEEEauthorrefmark{2}, Quan Z. Sheng\IEEEauthorrefmark{3}, Song Guo\IEEEauthorrefmark{4}}
% \IEEEauthorblockA{\IEEEauthorrefmark{2}College of
% Electronic and Information Engineering, Shenzhen University, Shenzhen, China\\
% \IEEEauthorrefmark{3}School of Computing, Macquarie University, Sydney, Australia\\
% \IEEEauthorrefmark{4}Department of Computer Science and Engineering, The Hong Kong University of Science and Technology, Hong Kong\\
% Email: zhaochonghe\_szu@163.com, yipeng.zhou@mq.edu.au, \{zsl, ttwang\}@szu.edu.cn\\ michael.sheng@mq.edu.au, songguo@cse.ust.hk}
% }
\DeclareRobustCommand*{\IEEEauthorrefmark}[1]{%
    \raisebox{0pt}[0pt][0pt]{\textsuperscript{\footnotesize\ensuremath{#1}}}}
% \author{\IEEEauthorblockN{Chonghe Zhao\IEEEauthorrefmark{1,2}, Yipeng Zhou\IEEEauthorrefmark{2}, Shengli Zhang\IEEEauthorrefmark{1}, Taotao Wang\IEEEauthorrefmark{1}, Quan Z. Sheng\IEEEauthorrefmark{2}, Song Guo\IEEEauthorrefmark{3}}
% \IEEEauthorblockA{\IEEEauthorrefmark{1}College of
% Electronic and Information Engineering, Shenzhen University, Shenzhen, China\\
% \IEEEauthorrefmark{2}School of Computing, Macquarie University, Sydney, Australia\\
% \IEEEauthorrefmark{3}Department of Computer Science and Engineering, The Hong Kong University of Science and Technology, Hong Kong\\
% Email: zhaochonghe\_szu@163.com, yipeng.zhou@mq.edu.au, \{zsl, ttwang\}@szu.edu.cn\\ michael.sheng@mq.edu.au, songguo@cse.ust.hk}
% }

%\IEEEauthorrefmark{1}Corresponding author: Shengli Zhang

\maketitle

\begin{abstract}
Proof of Authority (PoA) plays a pivotal role in blockchains for reaching consensus. % on decentralized nodes. 
%Clique is the most popular implementation of PoA due to its low requirement in communication overhead and energy consumption, which selects consensus nodes to generate blocks with a pre-determined order.% 
Clique, which selects consensus nodes to generate blocks with a pre-determined order, is the most popular implementation of PoA 
due to its low communication overhead and energy consumption.
However, our study unveils that the speed to process transactions by Clique is severely restricted by  1) the long communication delay of full blocks (each containing a certain number of transactions) between consensus nodes; and 2) occurrences of no-turn blocks, generated by no-turn nodes if an in-turn block generation fails. %e processed and broadcasted in time. 
Consequently, Clique struggles to support distributed applications requiring a high transaction processing speed, \emph{e.g.}, online gaming. To overcome this deficiency, we propose an Express Clique (ExClique) algorithm by improving Clique from two perspectives: compacting blocks for broadcasting to shorten communication delay and prohibiting the occurrences of no-turn blocks. For performance evaluation, we implement ExClique by modifying Geth of Ethereum, the software implementing Clique, and deploy a permissioned blockchain network by using container technology. The experimental results show that ExClique achieves a substantial enhancement in transactions per second (TPS). Specifically, it boosts TPS by 2.25$\times$ in a typical network with 21 consensus nodes and an impressive 7.01$\times$ in a large-scale network with 101 consensus nodes when compared to Clique.

\end{abstract}


\section{Introduction}\label{intro}
Blockchain is a distributed ledger system in which participants collectively share information in a decentralized and secure manner. It was first proposed in Bitcoin for decentralized asset transfer \cite{bitcoin} and further extended in Ethereum for decentralized applications  by introducing smart contracts \cite{ethereum}. Due to its notable characteristics in decentralization, immutability, and anonymity, blockchain has become a disruptive technology in the fields of Web 3.0 \cite{web3,web3_account}, Decentralized Finance \cite{mev},  Non-Fungible Token \cite{nft}, IoT \cite{IoT}, and Metaverse \cite{duan2021metaverse}. 
%{\bf YP: We study permission blockchains. Can we say the same thing for permission blockchain? If not, please add something for permission blockchains}

Since blockchain is a fully distributed system, consensus algorithms largely determine its security and efficiency. The consensus algorithm in blockchain plays a role in maintaining a consistent ledger among multiple nodes. To date, various consensus algorithms have been proposed for blockchains \cite{weak_consensus,pow_age,consensus_survey,consensus_1}, including Proof-of-Work (PoW), Proof-of-Stake (PoS), Practical Byzantine Fault Tolerant (PBFT), and Proof-of-Authority (PoA). Among them, PoA, which selects a set of authorized nodes to produce blocks in turn, gradually stands out as the dominant one due to its enhanced fault tolerance, reduced communication overhead, and energy efficiency \cite{comparative}.

%, as it bolsters fault tolerance, minimizes communication overhead, and is energy-efficient \cite{comparative}.

Clique \cite{clique_EIP225}, as a mainstream implementation of PoA, has been widely adopted in enterprises, such as Huobi, Binance, and Amazon \cite{heco,BSC,amazon}. In Clique, participating nodes follow a pre-determined order to process transactions and generate blocks. According to this order, an in-turn (consensus) node is responsible for generating the in-turn block within a specified time interval, \emph{i.e.}, a step. To avoid the single point failure of any in-turn node, about 50\%  consensus nodes act as no-turn nodes trying to generate no-turn blocks to process transactions if they fail to receive and verify the in-turn block by waiting for a random period of time \cite{clones_attack,unfairness_exploring}. Thanks to this simple yet effective design, Clique is incredibly cheap to run and maintain, contributing to its widespread popularity \cite{Clique_Adv}.

Despite its prevalence, %Clique has received tremendous research attention. However, 
the optimization of TPS (transactions per second) in Clique has received little attention in existing works, though TPS is one of the most important metrics to gauge the capability of blockchains \cite{fabric,privateBlockchain}. 
Sondhi et al. \cite{chaos_engineer} applied chaos engineering to demonstrate the better ability of  Clique to handle load than PBFT and Raft while working in a faulty environment. The work in \cite{performance_evaluation_blockchain} evaluated transaction confirmation time and TPS performance between different consensus algorithms to verify the suitability of  Clique for blockchain-based auditing systems. Islam et al. \cite{comparative} pointed out that fork is a main factor impacting TPS in Clique without proposing a solution. 
Due to the absence of TPS optimization in Clique, current blockchains adopting Clique still face the challenge of satisfying applications with massive transactions, \emph{e.g.}, online gaming and payment \cite{oneline_game,permission3}.

In this work, to optimize  TPS in Clique, we conduct a comprehensive analysis of  Clique. Our study reveals that the transaction processing speed of Clique is restricted by the long communication delay of full block broadcasting and occurrences of no-turn blocks. On the one hand, a Clique block containing a certain number of transactions should be broadcast to all consensus nodes for confirmation. Thus, a long broadcasting delay restricts Clique's capability to generate blocks.  On the other hand, in a given time interval,  an in-turn block should be generated by the in-turn node for others to accept this block. Yet, any failure to produce and transmit this block in time can lead to the creation of no-turn blocks by other nodes which prolongs block processing time and even results in empty blocks. These issues are detrimental to the TPS performance of blockchains adopting Clique.

To address these challenges, we propose the Express Clique (ExClique) algorithm, which can significantly improve  TPS from two perspectives. \emph{First}, we propose a novel compact block broadcast protocol, which can shorten the broadcasting delay of blocks by averting the transmission of redundant information. \emph{Second}, we explore the underlying reason for generating no-turn blocks, based on which we design two means to prohibit no-turn blocks. Specifically, we propose a tighter delay range and a differential order (in lieu of the pre-determined order) to dynamically designate in-turn nodes. The tighter delay range optimizes the random waiting time of no-turn nodes, reducing occurrences of forks and no-turn blocks in Clique. The differential order to configure in-turn nodes effectively prevents the ripple effect of no-turn nodes.\footnote{Once a no-turn block emerges, the pre-determined order of in-turn nodes will be disrupted, yielding consecutive no-turn blocks in subsequent time intervals. This is called the ripple effect in Clique.} Besides, a smart contract is introduced to eliminate unfairness among consensus nodes when  using the differential order.%, while lowering the risk of a successful \emph{Selfish Collusion Attack}.

%To address these challenges and bolster the TPS of blockchains adopting Clique, we propose the Express Clique (ExClique) algorithm, which can significantly improve the TPS performance from two perspectives. \emph{First}, we come up with a novel compact block broadcast protocol, which can shorten the broadcasting delay of blocks by averting the transmission of redundant information. \emph{Second}, we delve into the underlying causes of no-turn block generation and devise two schemes to prevent them. Specifically, we introduce an accurate delay range and a differential ordering for in-turn nodes. The former scheme optimizes the random waiting time of no-turn nodes, reducing forks and no-turn blocks within Clique. Additionally, when a no-turn block emerges, the pre-determined order of in-turn nodes can be disrupted, leading to consecutive no-turn block production in subsequent time intervals. ExClique effectively addresses this ripple effect by using the differential order to configure in-turn nodes dynamically.

To evaluate the performance of  ExClique, we implement it by modifying Geth of Ethereum and deploy  ExClique on a permissioned blockchain network by using container technology. The experimental findings unequivocally demonstrate the exceptional performance of ExClique. In a typical blockchain network comprising 21 consensus nodes, ExClique substantially shortens block broadcasting time by a factor of 5, remarkably mitigates fork rates by a factor of 12.4, and enhances TPS by a factor of 2.25. Significantly,  TPS gains are further amplified, reaching a remarkable 7.01$\times$ improvement in a larger blockchain network with  101 consensus nodes.

The rest of this paper is organized as follows. Section \ref{Related_Work} discusses related works. Section \ref{background} introduces the background of Clique. The TPS performance of Clique is analyzed in Section \ref{analysis_of_naive_clique}, which is further optimized in Section \ref{Design_Optimizing_Clique}. Section \ref{experiment} describes the experimental design and reports performance evaluation. Finally, Section \ref{conclusion} concludes this work.

\section{Related Work} \label{Related_Work}

%In this section, we discuss the related works from two perspectives: blockchain applications and TPS performance in Clique. % studies have delved extensively into its applications and TPS performance evaluation. 

% \subsection{Blockchain Applications}
% It has been investigated in \cite{permission1,permission2} that blockchains can be potentially adopted to support applications with a high requirement for privacy and security. For example, \cite{permission1} introduced a blockchain system designed for secure and efficient verification of academic records, enabling schools to transfer and validate student records securely upon request; \cite{permission5} explored how blockchain can be leveraged to address security issues in the Industrial Internet of Things (IIoT) such as data tampering, malicious control of devices, and privacy leakage.
% However, due to the low efficiency of consensus algorithms, such as Clique in permissioned blockchain, 
% blockchains confront the challenge of handling massive transactions in financial applications, \emph{e.g.},  online payment
%   \cite{permission3}. 
% %permissioned blockchains can only be deployed in a small-scale system. 

% \subsection{Clique Evaluation}\label{Related_Work}
%\color{blue}
Clique, as the most widely used consensus algorithm in blockchain,  has attracted huge attention from both academia and industry. However, existing works primarily explored the TPS performance of Clique through comparative evaluation.

%Toyodo et al. \cite{function_level} utilized a resource-profiling tool for Golang to dissect the function-level bottleneck of permissioned Ethereum blockchain, which transits from PoW to Clique. The experimental results underscored the imperative of optimizing the bottleneck function within Geth, the standard Ethereum software implementing Clique. 

In a comprehensive analysis of Clique's performance, Angelis \emph{et al.} \cite{CAP_analysis_1} first compared its performance with PBFT and Aura using the framework of the CAP Theorem. They highlighted that Clique achieves the shortest latency for committing a block among three algorithms, potentially resulting in high TPS  performance. Similarly, two other works evaluated Clique's performance by selecting PBFT as a baseline: 1) Sondhi \emph{et al.} \cite{chaos_engineer} utilized chaos engineering to assess various blockchain consensus algorithms, finding that Clique surpasses PBFT and Raft, while consistently maintaining favorable TPS even in faulty environments; 2) Ahmad \emph{et al.} \cite{performance_evaluation_blockchain} analyzed latency in transaction confirmation and TPS performance across different consensus algorithms, including PBFT, PoS, and Clique, within permissioned blockchains. This evaluation confirmed Clique's suitability for blockchain-based auditing systems. Additionally, like \cite{CAP_analysis_1} selecting Aura as a baseline, the works \cite{comparative,performance_evaluation_blockchain} conducted a comparative analysis between Clique and Aura to evaluate their security, TPS and block period. Their experimental findings revealed that Clique excels in availability and speed but lags in terms of security and consistency due to fork occurrences. These studies, focusing on Clique's performance evaluation, consistently highlight Clique's strength in achieving high TPS performance.
\color{black}

%Angelis et al. \cite{CAP_analysis_1} first compared its performance with PBFT and Aura using the framework of the CAP Theorem. They highlighted that Clique achieves the best latency for committing a block among the three algorithms, potentially resulting in high TPS (transactions per second) performance. Like \cite{CAP_analysis_1} selected PBFT as the baseline to compare the performance of Clique, Sondhi et al. \cite{chaos_engineer} harnessed chaos engineering to assess several blockchain consensus algorithms, which found that Clique  surpasses PBFT and Raft, while consistently maintaining favorable TPS even in faulty environments. Ahmad et al. \cite{performance_evaluation_blockchain} also scrutinized latency in transaction confirmation and the TPS performance across diverse consensus algorithms, including PBFT, POS, Clique, and so on, within permissioned blockchains. This evaluation confirmed Clique's suitability for blockchain-based auditing systems. Additionally, the works \cite{comparative,performance_evaluation_blockchain} conducted a comparative analysis between Clique and Aura to evaluate their security, TPS, and the block period. Their experimental findings revealed that  Clique excels in availability and speed but lags in terms of security and consistency due to fork occurrences. 
%prioritizes availability over consistency, leading to potential compromises.

%\begin{figure}[htpb]

The aforementioned studies shed light on the critical role of TPS in Clique. However, there are multiple potential factors impairing Clique's TPS, \emph{e.g.}, deadlocks. 
%Despite Clique's strengths,  occurrences of deadlocks significantly hamper its TPS performance, calling for solutions to mitigate this issue. 
The studies \cite{choice_ethereum_client, deadlock} reported that there is a random probability of generating two forked chains with the same weight, causing a deadlock in Clique, despite no malicious nodes trying to attack the network. Once a deadlock occurs, the blockchain systems adopting Clique fail to process transactions, and the TPS is thereby reduced to zero. To address this deficiency, Symons \emph{et al.} \cite{idchain} proposed a monitoring mechanism capable of detecting and reverting the system state when a deadlock arises.

Yet,   none of the existing works  systematically investigated all factors impairing TPS. Besides, they failed to propose specific schemes to boost Clique's TPS performance. These gaps will be filled by our work by proposing  ExClique. 
\color{black}

\section{ Background of Clique}\label{background}
%In this section, we introduce the background of Ethereum before we discuss the design of DEthna. 
%Before we introduce the design of DEthna, we provide the background of Ethereum in this section.
In this section, we introduce the basics of Clique and the process for generating in-turn and no-turn blocks by Clique. 
\subsection{Consensus Nodes}
In the initial phase of Clique, the system master authorizes $n$ consensus nodes (indexed from $1$ to $n$) to generate and verify blocks \cite{clique_EIP225,ethereum}. 
Normally, each block is generated in a fixed time interval, a.k.a., step, denoted by $t_b$. 
In each step, $n$ consensus nodes play three different roles as follows:
%The consensus nodes generate and verify blocks during the period of the step. Every step has a step duration time, i.e., block period $t_b+x$, where $t_b$ is a fixed value and specified in the genesis block, and $x$ is a random value. 
\begin{itemize}[leftmargin=10pt]
%\begin{itemize}
\item \textbf{\emph{In-turn Node}}: A particular consensus node is designated as the in-turn node that generates the in-turn block for this step with the highest priority. 
\item \textbf{\emph{No-turn Nodes}}: Multiple consensus nodes act as no-turn nodes which can generate the no-turn blocks with a lower priority if the in-turn consensus node fails.
\item \textbf{\emph{Forbidden Nodes}}:  The remaining consensus nodes are forbidden nodes that only receive and verify the blocks.
%in-turn block or no-turn blocks.  
\end{itemize}

Clique sets the number of consensus nodes for each type based on the following principle. To tolerate up to 
$\frac{n}{2}$ faulty nodes out of $n$ total consensus nodes, each consensus node can only generate a block (either in-turn block or no-turn block) within 
$\left\lfloor {\frac{n}{2}} \right\rfloor  + 1$ steps \cite{ethereum}. At Step $h$, Clique sets the consensus node with index $(h+1)\%n$  as the in-turn node, and  $\left\lfloor {\frac{{n + 1}}{2}} \right\rfloor - 1$ other consensus nodes (who do not generate any block within the past $\left\lfloor {\frac{n}{2}} \right\rfloor $ steps) as no-turn nodes. The rest of the consensus nodes are set as forbidden nodes. 
%there is a fixed in-turn node whose index $i$ is equal to $(h+1)\%n$ and 
%no-turn nodes 

\begin{figure}[t]
  \centering
  \includegraphics[width=\linewidth]{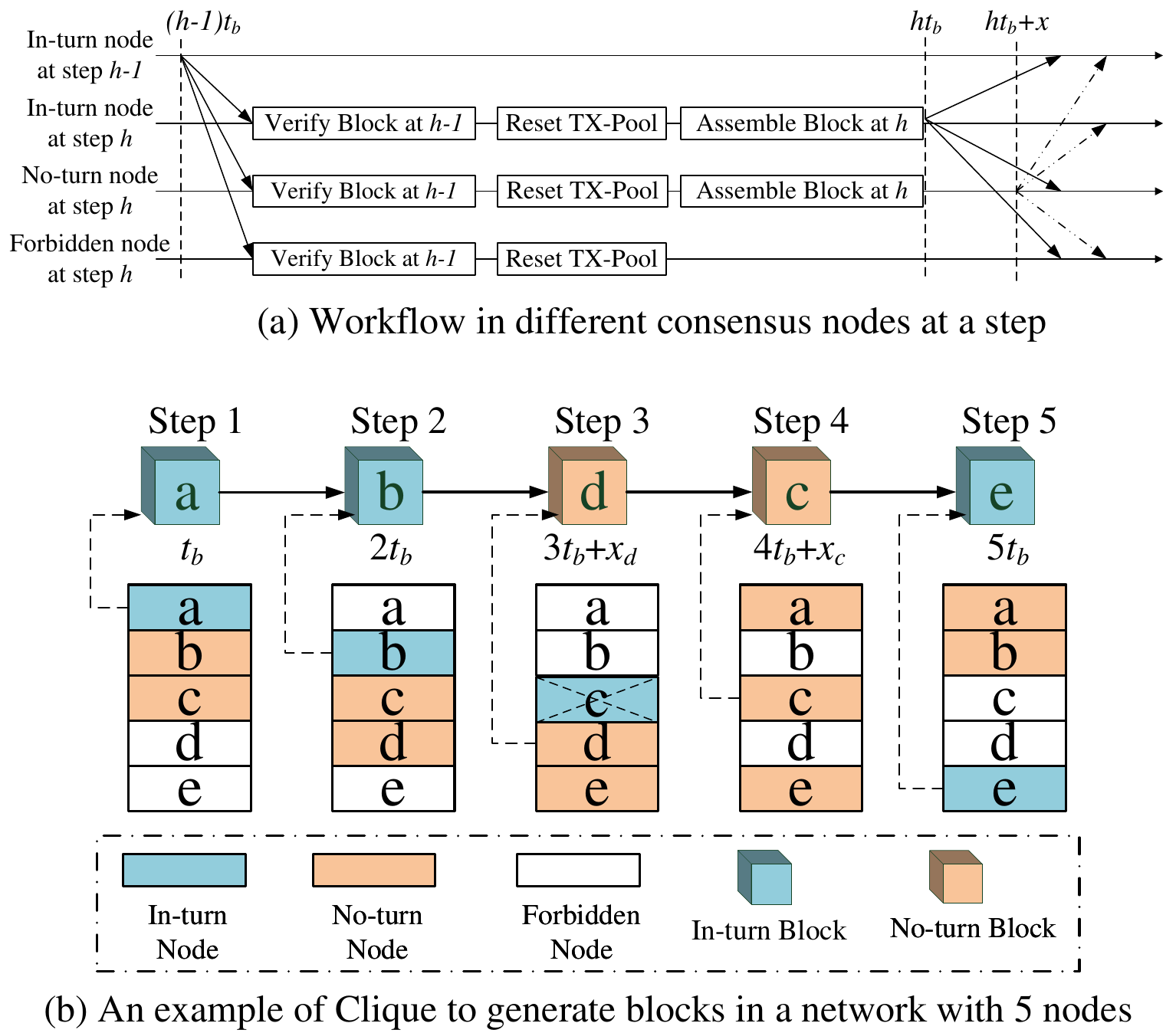}
  \caption{Specific Design of Clique: (a) Workflow in different consensus nodes; (2) An example to generate blocks.}
 % \Description{A woman and a girl in white dresses sit in an open car.}
  \label{workflow_clique1}
 %\vspace{-2pt}
\end{figure}

%\textbf{Three types of consensus nodes}: In each step, there are three types of consensus nodes to generate and verify blocks: in-turn consensus node that generates an in-turn block with high priority; no-turn consensus node that generates a no-turn block with low priority if the in-turn consensus node fails; forbidden node that only receives and verifies the in-turn block or the no-turn blocks. 
Suppose that all nodes can play their functions properly, the tasks conducted by each type of consensus node are presented in Fig.~\ref{workflow_clique1}(a). The in-turn node at Step $h-1$ broadcasts an in-turn block at time $(h-1)t_b$ to start next Step $h$. After receiving the in-turn block $h-1$, the consensus nodes perform different tasks according to their roles at Step $h$.  For the in-turn node, it verifies the in-turn block $h-1$, resets its transaction pool (TX-Pool) to remove invalid transactions, assembles a new in-turn block $h$ by selecting valid transactions from its TX-Pool, and broadcasts the new in-turn block $h$ at time $ht_b$ to kick off the next step.  For no-turn nodes, they verify the in-turn block $h-1$ and reset their TX-Pools.  For forbidden nodes, they only verify the in-turn or no-turn blocks and reset their TX-Pools. 
%, assemble a new no-turn block $h$, and broadcasts the new no-turn block $h$ if it does not receive and verify the in-turn block $h$ successfully until time $ht_b+x$;
%Note that, $x$ is a random delay within the range of $(0,w)$, where $w = (\left\lfloor {\frac{n}{2}} \right\rfloor  + 1)*500$ milliseconds.

%\textbf{Fixed order for in-turn nodes}: To tolerate up to $\frac{n}{2}$ faulty nodes among $n$ consensus nodes, each consensus node can only generate an in-turn or no-turn block within  $\left\lfloor {\frac{n}{2}} \right\rfloor  + 1$ steps \cite{ethereum}. Typically, at a specific step $h$, there is a fixed in-turn node whose index $i$ is equal to $(h+1)\%n$ and $\left\lfloor {\frac{{n + 1}}{2}} \right\rfloor - 1$ no-turn nodes who do not generate any in-turn or no-turn block within the past $\left\lfloor {\frac{n}{2}} \right\rfloor $ step.

\subsection{Block Generation}
However, the block generation process in practice is much more complicated than the case presented in Fig.~\ref{workflow_clique1}(a) because: 1) it is not guaranteed that all nodes can work reliably and 2) Clique is a purely distributed algorithm in which it is difficult to coordinate multiple nodes for block generation. Thus, it is common that no-turn blocks are generated, especially when Clique needs to process an excessive number of transactions.

To understand how no-turn blocks affect Clique, we present a concrete example in  Fig.~\ref{workflow_clique1}(b).
There are 5 consensus nodes in the system to generate blocks with the order from $a$ to $e$ at different steps. 
Blocks $a$ and $b$ are in-turn blocks generated by corresponding in-turn nodes (nodes $a$ and $b$) at $t_b$ and $2t_b$, respectively. %, because their generators (node 1 and node 2) are in-turn nodes at their step. 
At Step 3, the in-turn node $c$ fails to generate the in-turn block. % could not generate a block because a failure happened. 
In this case, one of two no-turn nodes (\emph{i.e.}, node $d$ and node $e$) will spontaneously generate no-turn blocks by waiting for a random period of time denoted by $x$. 
According to \cite{ethereum}, $x$ is randomly selected from the range $(0,w)$, where $w = (\left\lfloor {\frac{n}{2}} \right\rfloor  + 1)\cdot500$ milliseconds. Here, we assume that $x_d<x_e$, and then the no-turn block at Step 3 is generated by node $d$.
Moreover, there is a ripple effect for the influence of the no-turn block on the blockchain system.  If node $d$ generates the no-turn block at Step 3, it becomes a forbidden node (to meet the restriction that each  node can only generate a block in 
$\left\lfloor {\frac{n}{2}} \right\rfloor  + 1$ steps \cite{ethereum}) and cannot play the in-turn node role at Step 4 implying that a no-turn block will be generated again.

%who selected a minimum random delay $x_1$ waits until $3t_b+x_1$ to generate a no-turn block, and consider that no-turn node is node 4 to generate a no-turn block 4. Thereby, at step 4, node 4 becomes a forbidden node and node 3 becomes a no-turn node after it recovers from the failure. Similar to step 3, one of three no-turn nodes (node 1, node 3, and node 5) who selected a minimum random delay $x_2$ waits until $4t_b+x_2$ to generate a no-turn block and considers that no-turn node is node 3 to generate a no-turn block 3. Finally, node 5 waits until $5t_b$ to generate an in-turn block 5 at step 5. 

\section{TPS Analysis in Clique}\label{analysis_of_naive_clique}
%\section{Link Inference Model in Ethereum}\label{link}

In this section, we first analyze the factors influencing the TPS performance in Clique, and then define our problem to optimize TPS.

%conduct an in-depth analysis of the factors constraining TPS, ultimately leading to the identification of three key optimization directions guiding our design.

\subsection{TPS Formulation}\label{limitTPS}

As we have stated in the last section, a block is generated at each step. According to \cite{ethereumYellow}, each block contains a certain number of transactions, denoted as $m$, which is determined by setting a gas limit in the genesis block.
%each block contains a certain of $m$ transactions by setting a gasLimit in the genesis block.
Once $m$ is fixed, TPS is determined, which is no more than $\frac{m}{t_b}$ if there is no no-turn block generated. 

Nonetheless, as we have discussed in Fig.~\ref{workflow_clique1}(b), it is not uncommon that an in-turn node fails to generate an in-turn block resulting in the generation of no-turn blocks. Once no-turn blocks are generated, TPS is compromised from the two following aspects:
\begin{itemize}[leftmargin=10pt]
%\begin{itemize}
    \item The time interval of a step lasts longer than $t_b$ because no-turn nodes wait for additional $x$ time before generating blocks, where $x$ is randomly drawn from $(0, w)$. 
    \item If the previous step lasts much longer than $t_b$, the consensus nodes may not have sufficient time to generate full blocks in the current step. Instead, an empty block without any transaction is generated. 
\end{itemize}
%Based on our discussion, we can express the expected TPS as:

\begin{figure}[t]
  \centering
  \includegraphics[width=\linewidth]{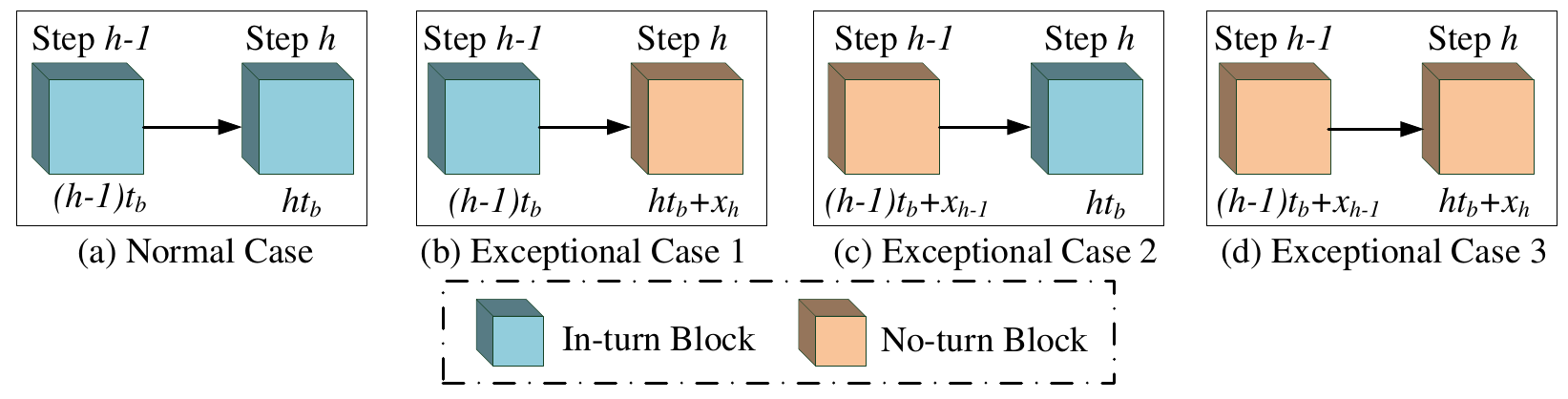}
  \caption{Four cases at a single step to affect the TPS performance in Clique.}
 % \Description{xxx.}
  \label{case}
  \vspace{-2pt}
\end{figure}

Based on our discussion, we derive the expected TPS as:
\begin{equation}\label{EQ:TPSDef}
\Lambda = p_{0} \cdot\Lambda_0 +(1-p_0) \cdot\Lambda_{0-}.
\end{equation}
Here, $p_0$ represents the probability that in-turn blocks are generated at each step, with TPS equal to $\Lambda_0=\frac{m}{t_b}$. $\Lambda_{0-}$ denotes the expected TPS when no-turn blocks are generated, which is very complicated for analysis.
Next, we analyze $\Lambda_0$ and $\Lambda_{0-}$ by modeling the block generation process with four cases, as shown in Fig.~\ref{case}. In this context, $\Lambda_{0-}$ is further decomposed into $\Lambda_1$, $\Lambda_2$ and $\Lambda_3$ corresponding to Exceptional Cases 1, 2, and 3, respectively, in Fig.~\ref{case}.

%In Clique, there is a normal case to achieve maximum TPS performance at a single step if two consecutive blocks committed into the global ledger are in-turn blocks. However, in practice, the presence of forks or other failures of in-turn nodes can lead to the inclusion of no-turn blocks, which can result in three exceptional cases of degraded TPS at a single step. The total TPS of Clique is determined jointly by these four cases, and these four cases at a single step, depicted in Fig.\ref{case}, are discussed as follows.

\noindent\textbf{\emph{Normal Case}}: If an in-turn block is generated at Step $h$ containing $m$ transactions, it implies that the in-turn node completes the following operations within $t_b$ time: 1) broadcasting the in-turn block to all nodes in $b_{h-1}(m)$ time; 2) verifying the in-turn block on all consensus nodes within $v_{h-1}(m)$ time; 3) resetting TX-Pool within $r_{h}(m)$ time; and 4)  assembling the block at step $h$ within $a_{h}(m)$ time. 
%As described in the last section, the in-turn node at step $h$ needs to finish the four tasks within a step duration $t_b$: receives the last in-turn block $h-1$, verifies the last in-turn block $h-1$, resets its local TX-Pool, and assemble the new in-turn block $h$. Thus, a blockchain system adopting Clique operates normally when the following condition is satisfied:
%In a word, the total time cost must be no more than the time interval of a step. Hence, we have 
That is, the total time must be no more than the time interval of a step. Hence, we have 
\begin{equation}\label{condition}
{f_h}(m) = {b_{h - 1}}(m) + {v_{h - 1}}(m) + {r_h}(m) + {a_h}(m) \le {t_b}.
\end{equation}
Here, $f_h(m)$  is the total time cost to generate the in-turn block and $m$ is a tuneable parameter. 

%number of transactions in a block; ${f_h}(m)$ represents the total time required for four tasks; $b_{h-1}(m)$ is the time to broadcast last in-turn block; $v_{h-1}(m)$ is the time to verify last in-turn block; $r_{h}(m)$ is the time required to reset TX-Pool; and $a_{h}(m)$ is the time to assemble block at step $h$. According to the analysis in \cite{bbp,hcb}, $b_{h-1}(m)$, $v_{h-1}(m)$, $r_{h}(m)$, and $a_{h}(m)$ all increase with $m$, the maximum TPS of the normal case is thereby achieved when $f_h(m)=t_b$, which is calculated by:
To improve TPS, we should set $m$ as large as possible. Thus, the maximum value of $m$ is 
\begin{equation}\label{idealTPS1}
m^* = \arg \max_m {f_h}(m)\leq t_b.
\end{equation}
It implies that the TPS in the normal case is limited by 
\begin{equation}\label{idealTPS2}
\Lambda_{0} = \frac{m^*}{t_b}.
%\frac{{\mathop {\arg \max {f_h}(m)}\limits_m }}{{h{t_b} - (h - 1){t_b}}}= \frac{{\mathop {\arg \max {f_h}(m)}\limits_m }}{t_b}
\end{equation}
% As shown in Fig.\ref{case}(a), the ideal case at step $h$ is that two consecutive blocks are the in-turn blocks, where the in-turn block $h-1$ is generated at time $(h-1)t_b$ and the in-turn block $h$ is generated at time $ht_b$. The maximum TPS achievable in this ideal case can be calculated as follows:
% \begin{equation}\label{idealTPS}
% TPS_{max} = \frac{{\mathop {\arg \max {f_h}(m)}\limits_m }}{{h{t_b} + {x_h} - (h - 1){t_b}}}= \frac{{\mathop {\arg \max {f_h}(m)}\limits_m }}{t_b}
% \end{equation}

\noindent\textbf{\emph{Exceptional Case 1}}: As shown in Fig.~\ref{case}(b), 
this exceptional case occurs when the in-turn node fails to generate an in-turn block at Step $h$ following an in-turn block at Step $h-1$. 
In this case,  $\left\lfloor {\frac{{n + 1}}{2}} \right\rfloor - 1$ no-turn nodes will compete to generate a no-turn block. Since the waiting time of each no-turn node obeys the uniform distribution $U(0, w)$, the expected minimum waiting time of these no-turn nodes ($x_h$) is the expected minimum value of  $\left\lfloor {\frac{{n + 1}}{2}} \right\rfloor - 1$ uniform random variables obeying $U(0, w)$. This waiting time $x_h$ is further denoted by
$\Delta_1 =\frac{w}{\left\lfloor {\frac{{n + 1}}{2}} \right\rfloor}$.
Thereby, $\Lambda_1 $ is 

%The no-turn block $h$ is generated after a random delay $x_h$ (i.e., broadcasted at time $ht_b+x_h$) and the in-turn block $h-1$ is generated at time $(h-1)t_b$. Since $x_h$ follows the uniform distribution of $(0,w)$, the average value of $x_h$ is $\frac{w}{2}$. The TPS performance, in this case, is  decreased from $TPS_{0}$ to:
\begin{equation}\label{tps1}
\Lambda_{1} = \frac{m^*}{t_b+\Delta_1}.
%\frac{{\mathop {\arg \max {f_h}(m)}\limits_m }}{{h{t_b} + {x_h} - (h - 1){t_b}}} = \frac{{\mathop {\arg \max {f_h}(m)}\limits_m }}{{{t_b} + {x_h}}}=\frac{{\mathop {\arg \max {f_h}(m)}\limits_m }}{{{t_b} + {\frac{w}{2}}}}
\end{equation}
Given that  $t_b$ and $w$ are usually set to $t_b=3$ seconds and $w = (\left\lfloor {\frac{n}{2}} \right\rfloor  + 1)\cdot500$ milliseconds \cite{ethereum},  we have $\Delta_1\approx 500$ milliseconds and the occurrence of Exceptional Case 1 can roughly lower TPS by 14.3\%.

%by 16.7\%. 

 \noindent\textbf{\emph{Exceptional Case 2}}: As shown in Fig.~\ref{case}(c), Exceptional Case 2 occurs when an in-turn block $h$ is generated  following a no-turn block generated at Step $h-1$. As we have stated, when a no-turn block is generated at Step $h-1$, the duration of Step $h-1$ lasts longer than $t_b$. However, Step $h$ will be over at time $ht_b$ exactly, which implies that the in-turn node at Step $h$ does not have sufficient time to process $m$ transactions (\emph{i.e.}, $ht_b-(h-1)t_b-x_{h-1}<t_b$), and hence an empty in-turn block is generated without any transaction. Thus, we have $\Lambda_2 =0$. 
 
% In this case, the no-turn block $h-1$ is generated after a random delay $x_{h-1}$ (i.e., broadcasted at time $(h-1)t_b+x_{h-1})$ and the in-turn block $h$ is generated at time $ht_b$, i.e., the step duration time is $t_b-x_{h-1}$. According to eq.(\ref{condition}), it takes $t_b$ for an in-turn block with 
%${\mathop {\arg \max {f_h}(m)}\limits_m }$ transactions to be generated, the shorter step duration time, in this case, would cause the generation of an empty in-turn block without any transactions since the in-turn node would first assemble and broadcast an empty block at time $ht_b$ when the in-turn block with ${\mathop {\arg \max {f_h}(m)}\limits_m }$   transactions cannot be assembled in time at time $ht_b$ \cite{ethereum,hcb}. Therefore, in this exceptional case, the TPS performance is decreased to:
%\begin{equation}\label{tps2}
%TP{S_2} = \frac{0}{{{t_b} - {x_{h - 1}}}}TPS_{2} = 0\end{equation}

\noindent\textbf{\emph{Exceptional Case 3}}: As shown in Fig.~\ref{case}(d), Exceptional Case 3 occurs when two consecutive blocks are generated by no-turn nodes. Let $\delta_{h-1}$ and $\delta_{h}$ denote the additional waiting times at Step $h-1$ and Step $h$, respectively (\emph{i.e.}, $\delta_{h-1} = x_{h-1}$ and $\delta_{h} = x_h$). These are random variables that follow the same distribution, specifically the minimum waiting time of $\left\lfloor \frac{{n + 1}}{2} \right\rfloor - 1$ uniform random variables distributed as $U(0, w)$.
%Let $\delta_{h-1}$ and $\delta_{h}$ denote the additional waiting time at Step $h-1$ and Step $h$, respectively, which are random variables obeying the same distribution, \emph{i.e.}, the minimum waiting time of $\left\lfloor {\frac{{n + 1}}{2}} \right\rfloor - 1$ uniform random variables obeying $U(0, w)$. 
Exceptional Case 3 can be analyzed with two separate scenarios: $\delta_h\geq \delta_{h-1}$ and $\delta_h<\delta_{h-1} $. 

If  $\delta_h\geq \delta_{h-1}$, it implies that no-turn nodes at Step $h$ have sufficient time to process $m$ transactions. TPS in this scenario can be  expressed as
\begin{equation}\label{tps31}
\Lambda_{31} =  \frac{m^*}{t_b+ \mathbb{E}[\delta_h-\delta_{h-1}|\delta_h\geq \delta_{h-1}]}.
\end{equation}
Here, $\mathbb{E}$ denotes expectation. 

If  $\delta_h<\delta_{h-1}$, it implies that no-turn nodes at Step $h$ do not have enough time to process $m^*$ transactions, implying that $\Lambda_{32} =0$, similar to Exceptional Case 2.

Since $\delta_h$ and $\delta_{h-1}$ have an identical distribution, the odds of each scenario is 50\%. Thus, $\Lambda_3 = \frac{1}{2}\cdot \frac{m^*}{t_b+ \mathbb{E}[\delta_h-\delta_{h-1}|\delta_h\geq \delta_{h-1}]}.$

%, where the no-turn block $h-1$ is generated at time $(h-1)t_b+x_{h-1}$ and the in-turn block $h$ is generated at time $ht_b+x_h$. In this exceptional case, the TPS performance is determined by $x_{h-1}$ and $x_h$: it can be regarded as exceptional case 2 to generate an empty block when $x_h-x_{h-1}<0$;   exceptional case 1 when $x_h-x_{h-1}>0$.  The specific TPS in this case can be calculated as follows:

%\begin{equation}\label{tps3}
%T P S_3= \begin{cases}\frac{{\mathop {\arg \max {f_h}(m)}\limits_m }}{t_b+x_h-x_{h-1}} \approx T P S_1, & x_h-x_{h-1}>0 \\ \frac{0}{t_b+x_h-x_{h-1}} =T P S_2, & x_h-x_{h-1}<0\end{cases}
%\end{equation}

%When $x_h-x_{h-1}>0$, the average value of $x_h-x_{h-1}$ is

\subsection{Problem Statement}

By wrapping up the TPS performance of all cases, the expected TPS of Clique is presented as follows:
%Let $p_0$ denote the probability of the normal case occurring, $p_1$ represent the probability of exceptional case 1 occurring, $p_2$ signify the probability of exceptional case 2 taking place, and $p_3$ correspond to the probability of exceptional case 3 unfolding. Combining with eq.(\ref{idealTPS})-(\ref{tps3}), the total TPS of Clique is thereby calculated as:
\begin{equation}\label{totalTPS}
\begin{aligned}
\Lambda &= {p_0}\cdot\Lambda_0  + {p_1}\cdot\Lambda_1+ {p_2}\cdot\Lambda_2 + {p_3}\cdot\Lambda_3 \\
&= {p_0}\cdot\Lambda_0 + {p_1}\cdot\frac{{\Lambda_0}}{{1 + \frac{{{\Delta_1}}}{{{t_b}}}}}  + \frac{p_3}{2}\cdot \frac{\Lambda_0}{1+ \frac{\mathbb{E}[\delta_h-\delta_{h-1}|\delta_h\geq \delta_{h-1}]}{t_b}}.
\end{aligned}
\end{equation}

%\begin{equation}\label{totalTPS}
%\begin{aligned}
%TP{S_t} = {p_0}*TP{S_0} + {p_1}*TP{S_1} + {p_2}*TP{S_2} + {p_3}*TP{S_3} \\= {p_0}*TP{S_0} + {p_1}*\frac{{TP{S_0}}}{{1 + \frac{{{x_h}}}{{{t_b}}}}} + {p_2}*0 + (\frac{{{p_3}}}{2}*\frac{{TP{S_0}}}{{1 + \frac{{{x_h}}}{{{t_b}}}}} + \frac{{{p_3}}}{2}*0)
%\end{aligned}
%\end{equation}

\begin{figure}[t]
  \centering
  \includegraphics[scale=0.40]{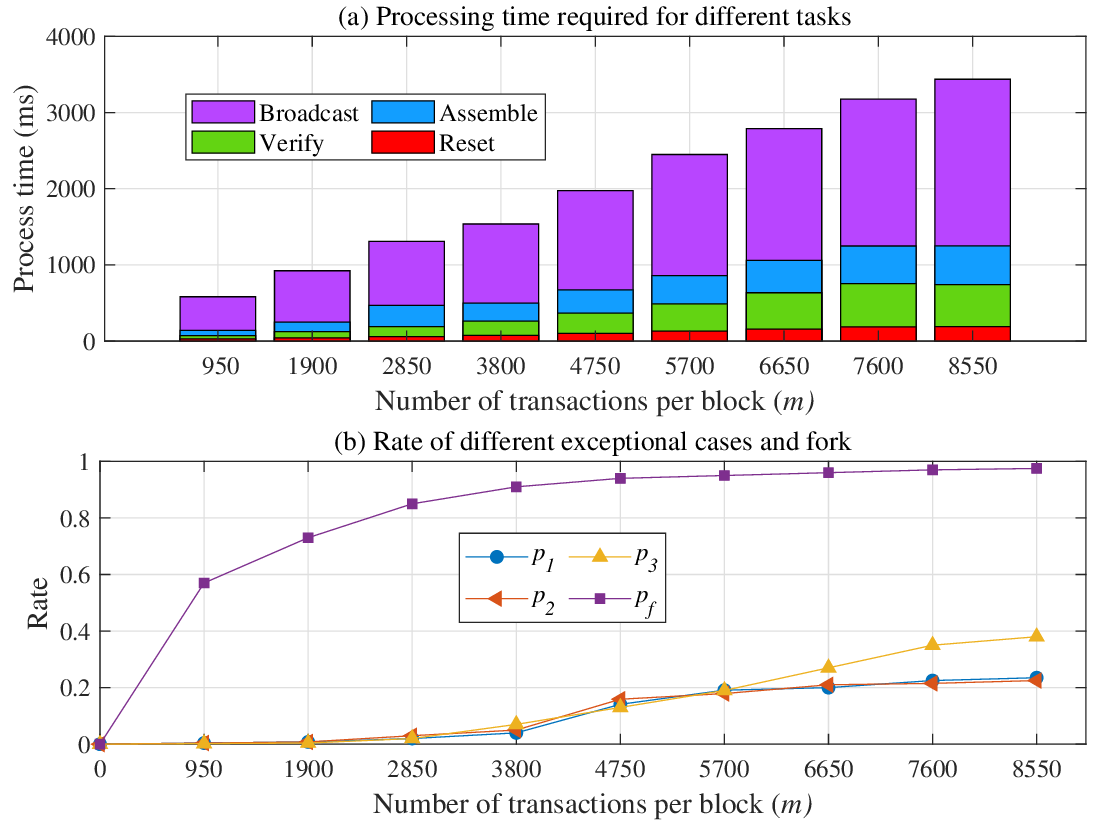}
  \caption{Investigation Results on Clique: (a) Processing time required for each task; (b) Rate of different exceptional cases and fork.}%rate of forks and different exceptional cases.}
  %\Description{A woman and a girl in white dresses sit in an open car.}
  \label{process}
%\vspace{-12pt}
\end{figure}

It is easy to verify that $\Lambda_1, \Lambda_2, \Lambda_3<\Lambda_0$. 
From Eq.~\eqref{totalTPS}, we can see that $\Lambda$ can be optimized by maximizing $\Lambda_0$ and minimizing $p_1, p_2$ and $p_3$ (equivalent to maxmizing $p_0$). %there are two optimizing directions to improve the total TPS of Clique: 1) Improve TPS of the normal case (i.e., improve $TPS_0$); 2) Reduce the probability of each exceptional case (i.e., $p_1$, $p_2$, and $p_3$).    
In the next section, we will explore how to maximize $\Lambda$ in ExClique.

\section{Optimizing TPS in ExClique}\label{Design_Optimizing_Clique}
In this section, we introduce the design of the ExClique algorithm in order to optimize TPS, including a proactive compact block (PCB) protocol to maximize  $\Lambda_0$, an accurate delay range to minimize $p_1$ and $p_2$, and a differential order of in-turn nodes to minimize $p_3$. 

\subsection{Maximizing $\Lambda_0$}\label{tcb}
According to Eqs.~(\ref{condition}) and~(\ref{idealTPS2}), the TPS of the normal case, denoted as $\Lambda_0$, is determined by the total time cost of four tasks: \emph{block broadcasting}, \emph{block verification}, \emph{TX-Pool reset}, and \emph{block assembly}. We improve $\Lambda_0$ by 1) identifying the block broadcasting task as the most critical one dominating the total time cost and 2) developing a PCB protocol to shrink communication traffic so as to diminish the total time cost.

\noindent\textbf{\emph{Main factor dominating $\Lambda_0$}}:  %In real-world scenarios, there often exists significant variance in the time required for these different tasks. To attain a more comprehensive insight into these tasks and 
To identify the main factor that constrains $\Lambda_0$, we conduct the experiment with a permissioned blockchain network comprising 21 consensus nodes executing Clique.  In this experiment, we set the step duration time $t_b=3$ seconds. 
To test the TPS performance, we vary the number of transactions contained in each block from 950 to 8,550.\footnote{Other parameters of this experiment are the same as those in  Section \ref{setup}. For more detailed parameter settings, please refer to Section \ref{setup}.}
In Fig.~\ref{process}(a), we present our experimental results with the x-axis representing the number of transactions per block and the y-axis representing the time cost of each block. In addition, we decompose the time cost into four parts corresponding to the time cost of each task in Eq.~\eqref{condition}.

%We compared the specific time required for each task under different values of $m$ and the experimental results are illustrated 
  %Here, the horizontal axis represents the number of transactions in a block, denoted as $m$. 
From the experimental result, we can observe that the total time cost and the time cost of each task rise with respect to the increase of  $m$, and notably, the time cost of block broadcasting dominates the total time cost. 
%From Fig. \ref{process}, it is evident that the time dedicated to block broadcasting constitutes the most significant portion of the overall time required for the four tasks. 
\begin{figure}[t]
  \centering
 \includegraphics[scale=0.46]{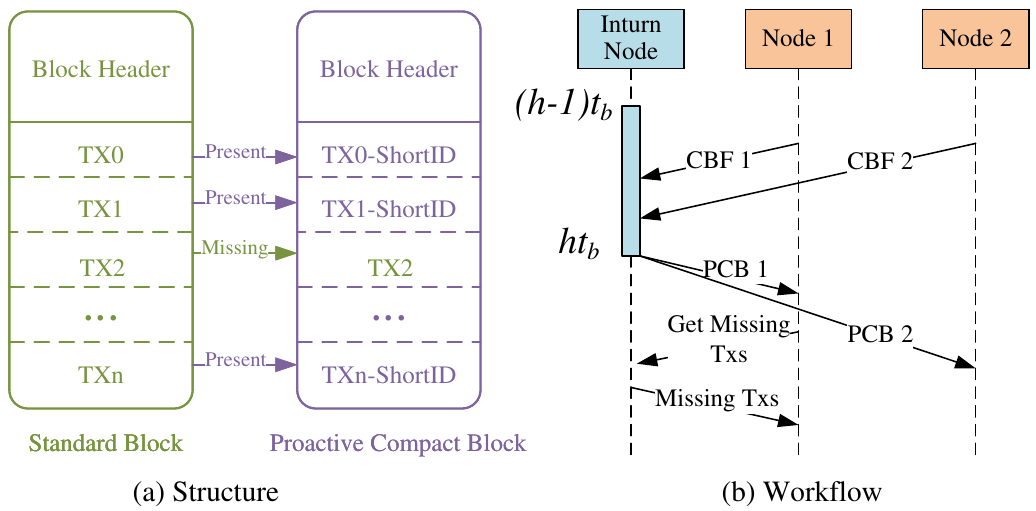}
 % \caption{Detailed design of filter-based compact block: (a) structure; (b) workflow.}
   \caption{Detailed Design of PCB: (a) structure; (b) workflow.}

 % \Description{A woman and a girl in white dresses sit in an open car.}
  \label{tcb_protocol}
%\vspace{-13pt}
\end{figure}
This issue arises from the rapid increase in standard block size driven by the surge in the number of transactions \cite{hcb}. Therefore, $\Lambda_0$ can be substantially improved if we can effectively reduce the time cost of block broadcasting. 

%The reason is rooted in the rapid expansion in the size of a standard block because of the surge in the number of transactions within a block \cite{hcb}. Therefore, $\Lambda_0$ can be substantially improved if we can effectively reduce the time cost of block broadcasting. 

\noindent\textbf{\emph{PCB for speeding up block broadcasting}}: 
To speed up block broadcasting, we design a proactive compact block (PCB) protocol. The design of PCB is based on the fact that the consensus nodes frequently exchange transactions with each other which actually already exist in their TX-Pools. We use a counting bloom filter (CBF) \cite{cbfCaclulate} to indicate the missing status of each transaction on each consensus node. By exchanging CBFs prior to transaction broadcasting, the consensus nodes can avoid broadcasting those redundant transactions which are already owned by receiving nodes.

%compact block \cite{compact} and Counting Bloom Filter . We first briefly review the basic compact block. Basic 
It has been investigated in Bitcoin that nodes exchange transactions in their TX-Pools before these transactions are packaged in a block, and thus the similarity of their TX-Pools is very high \cite{similarity}. To improve the block broadcasting speed in Bitcoin, a compact block protocol was proposed \cite{compact_1,compact}, 
%compact block, which compresses block by obviating the need to broadcast the transactions already in the receiving node’s TX-Pool, is a good way to speed up block broadcasting and adopted in Bitcoin. 
in which transactions in a standard block are represented by their hashes. Receiving nodes can identify these transactions from their local TX-Pools by these hashes. 

Unfortunately, it is infeasible to straightly apply the compact block protocol in blockchain systems adopting Clique in which the period to generate a block is much shorter. % requires a short period to generate a block, because the short period 
As a result, nodes in Clique have a very low similarity of TX-Pools, which can excessively incur additional communication rounds for broadcasting missing transactions \cite{hcb}. % missing from TX-Pools of receiving nodes \cite{hcb}. %In cases where certain transactions are missing and not stored in their local TX-Pools, an additional communication round is required to obtain these missing transactions. Regrettably, 
%There is a large probability of incurring additional communication rounds, 
For example, the probability of incurring additional communication rounds exceeds 90\% in Ethereum Mainnet. %, in which most nodes adopt the same software Geth as the consensus nodes in Clique, 
It implies that the efficiency of the compact block protocol will significantly deteriorate in blockchains adopting Clique \cite{bbp,hcb}.

Since the in-turn node and the time allocated for generating in-turn blocks at each step can be easily identified in blockchains adopting the PoA consensus algorithm (whether Clique with a fixed order or our ExClique with a differential order, as discussed in Section \ref{relationshiptpsapple}), we design PCB by introducing CBF which avoids incurring extra communication rounds in most cases.   As illustrated in Fig.~\ref{tcb_protocol}(a), each transaction within a PCB block is represented as either ``missing" or ``present" based on whether it exists in the receiving node's TX-Pool or not. If a transaction is tagged as ``missing", the complete transaction data will be included in the PCB block without the need for additional communication rounds. Conversely, for transactions tagged as ``present", only a short hash (a 6-byte short ID, as depicted in Fig.~\ref{tcb_protocol}(a)) is embedded into the PCB block to minimize broadcasting overload.

To tag transactions, each consensus node maintains a CBF. This data structure allows a consensus node to represent all transactions within its local TX-Pool concisely, enabling the instant detection of missing transactions. Besides, unlike the traditional Bloom Filter that can only support the \emph{add} operation, CBF  supports both \emph{add} and \emph{remove} operations so that it can be updated dynamically \cite{bloom_survey}. In other words, an element can be added  or removed when a new transaction is stored or an invalid transaction is removed in TX-Pool, % or removed when an invalid transaction is removed from TX-Pool, 
\emph{e.g.}, the transaction that has been packaged in a block.
%{\bf YP: no idea why a packaged transaction is invalid. } 
Due to this dynamic maintenance feature, each consensus node can easily create and maintain a CBF.

The detailed workflow of PCB is presented in Fig.~\ref{tcb_protocol}(b). At Step $h$, before the birth of the in-turn block in this step, %generates the in-turn block, 
node 1 and node 2 (either forbidden or no-turn nodes) send their respective CBFs to the in-turn node. The in-turn node, through the examination of their CBFs, identifies missing and present transactions for node 1 and node 2, respectively. Then, the in-turn node can tailor different PCB blocks (PCB 1 and PCB 2) for them. These PCB blocks are then broadcasted, with PCB 1 being sent to node 1 and PCB 2 to node 2. Upon the reception of their respective PCB blocks, node 1 and node 2 can reconstruct the standard block by retrieving transactions from their TX-Pools.

Although the chance is rare, it is worth noting that using CBF in ExClique may also consume additional communication rounds for broadcasting missing transactions. First, CBF cannot guarantee the uniqueness of the hash ID for each transaction, \emph{i.e.}, the false negative problem in CBF \cite{false_negative_problem}.
If hash IDs conflict, CBF cannot exactly tell missing and present transactions in receiving nodes' TX-Pools. Second, transactions in the blockchain are dynamic. It will consume additional communication rounds if CBF cannot factor in all changes of transactions in time, \emph{e.g.,} removal of a transaction in  TX-Pools after sending CBF but before receiving PCB.

%and alterations in the transaction pool \cite{hcb}, there exists a small possibility that node 1 or node 2  fails to reconstruct the standard block from their TX-Pools. In such cases, some missing transactions are tagged as the present transactions so that the in-turn node does not embed them into the FCB, they need to be requested from the in-turn node. Specially, for the false negative problem of CBF, a missing transaction and a present transaction share the same location in the CBF due to the limited space of CBF; for the alterations in the transaction pool, a missing transaction is stored in the TX-Pool of node1 or node2 but is removed when the FCB arrives due to the limited space of TX-Pool. To mitigate these two limitations, we also select a reasonable space for CBF and add a secondary Pool by referring to \cite{CountingBloomFilter,hcb}. }

%{\bf YP: can you better explain the false negative problem  and   alterations in the transaction pool, and explain why receiving nodes fail to reconstruct blocks}

\noindent\textbf{\emph{Compression rate of PCB}}: 
Intuitively speaking, how much time cost can be reduced by PCB is dependent on the number of short transactions that can be represented by hash values (6-byte short IDs). In contrast, the smallest traffic consumed by a full transaction for token transfer \cite{transaction_size}  is around 110 Bytes. It implies that the compression rate is at least $110/6\approx 18$ times by replacing a full transaction with a short ID in PCB.

%As discussed in \cite{hcb}, block broadcasting time in the compact block-like protocol is determined by communication rounds and the size of the block. In our tailored compact block, the communication rounds are minimized by including some missing transactions. Thus, compared to a standard block, the benefit of our tailored compact block can be reflected in its size. 
%For a standard block consisting of a header and $m$ full transactions, its size is calculated by 
%\begin{equation}
%{s_{sb}} = {s_h} + m{s_t}
%\end{equation}
%where $s_h$ is the size of header, and $s_t$ is the size of a transaction. For our tailored compact block with transaction short IDs and some full transactions, its size is calculated by 
%\begin{equation}
%   {s_{tcb}} = {s_h} + (1-\alpha) m{s_i} +  \alpha m{s_t}
%\end{equation}
%where $\alpha$ is the proportion of full transactions included in a tailored compact block, and $s_i$ is the size of a transaction short ID (i.e., 6 Bytes). Thus, the difference between $s_{sb}$ and $s_{tcb}$ is 
%\begin{equation}\label{difference}
%   {s_{sb}- s_{tcb}} =   m(1-\alpha) ({s_t}-{s_i})
%\end{equation}
%According to \ref{difference}, the bandwidth that can be saved by our tailored compact block is determined by $\alpha$ and $s_t$ for a fixed $m$. 
The fraction of short transactions is related to the network size. In the Ethereum MainNet with thousands of nodes \cite{hcb},
this fraction is around 0.9. For the blockchain network adopting Clique, the network has only dozens of nodes and the fraction of short transactions exceeds  0.9.  Therefore, our PCB protocol can considerably shrink communication traffic, and hence diminish the time cost of block broadcasting.
%  This phenomenon arises because the size of the standard block, which contains all full transactions, increases significantly with $m$, resulting in substantial broadcasting delays. Consequently, we arrive at our first optimization direction for Clique:
%\textbf{Optimizing Direction 1}: To enhance the TPS performance of the normal case, an effective approach is to expedite block broadcasting by reducing the size of the block used in broadcasting.
%According to eq.(\ref{totalTPS}), mitigating the probability of each exceptional case ($p_1$, $p_2$, and $p_3$) is imperative to improve total TPS.  
\subsection{Minimizing $p_1$ and $p_2$}
We minimize $p_1$ and $p_2$ by using the minimization of the fork probability in Clique as the surrogate objective because forks are caused by the generation of no-turn blocks which then trigger Exceptional Cases 1 and 2. 
Clique adopts the parallel block generation mechanism to effectively prevent the single-point failure of in-turn nodes from havoc on the entire network. Nonetheless, the cost of the parallel block generation mechanism lies in the occurrence of forks. %where more than one no-turn node proposes a block at the same step. 
Typically, if the in-turn block cannot be generated in time, no-turn nodes will try to generate no-turn blocks, which potentially causes forks. Once forks occur, it is possible that Exceptional Cases 1 and 2 described in the last section will occur. 
%with high priority according to the block selection rule \cite{ghost_protocol}. However, the no-turn block would be selected to trigger an exceptional case 1 if the forks occur frequently.

 \noindent\textbf{\emph{Forks influencing $p_1$ and $p_2$}}: 
 To demonstrate the relationship between forks and the two probabilities ($p_1$ and $p_2$), we measure the fork event probability (denoted by $p_f$) together with $p_1$ and $p_2$ in the experiment presented in Fig.~\ref{process}(a). Then, we plot $p_f$, $p_1$ and $p_2$ v.s. the number of transactions  in each block, \emph{i.e.}, $m$,  in Fig.~\ref{process}(b). 
 %Meanwhile, an exceptional case 2 would be triggered when an in-turn block is selected after the no-turn block in exceptional 1. A clear relationship between $p_1$, $p_2$, and forks rate $p_f$ is shown in Fig. \ref{process} (b) by using the same permissioned blockchain network in Section \ref{tcb}. 
 
 The experimental results show that: 1) with the increase of $m$ in each block, the fork probability $p_f$ rapidly rises; 2) both $p_1$ and $p_2$ have a strong positive correlation with $p_f$; 3) when $p_f>0.8$, we can see a significant increase of $p_1$ and $p_2$. Therefore, it is effective to use the minimization of $p_f$ as a surrogate objective for minimizing  $p_1$ and $p_2$. 

 %When forks occur, the consensus nodes would select the in-turn block or no-turn block according to the block selection rule \cite{ghost_protocol}. As shown in Fig.\ref{case} (b), exceptional case 1 occurs if forks cause a no-turn block to be selected and committed after an in-turn block. Exceptional case 2 will be triggered after an in-turn block is selected and committed after the no-turn block in exceptional case 1. A clear relationship between $p_1$, $p_2$ and forks $p_f$ is shown in Fig.\ref{process} (b) by using the same permissioned blockchain network in Section \ref{tcb}. The experimental results show that both $p_1$ and $p_2$ are positively related to $p_f$ when $p_f$ is larger than 0.9. 
 
%Obviously, the fork rate and the probability of no-turn block selected and committed into the global ledger are thereby positively related.

\noindent\textbf{\emph{Modeling fork probability}}: According to the description in the last section, the in-turn node at Step $h-1$ broadcasts the in-turn block at time $(h-1)t_b$, while a no-turn node also broadcasts its no-turn block if it cannot receive and verify the in-turn block successfully until time $(h-1)t_b+x$. In other words, a fork will occur at a step when the random delay $x$ is smaller than the total time spent on broadcasting and verifying the in-turn block. Since $x$ follows the uniform distribution $U(0,w)$, the probability of a fork incurred by a single no-turn node at Step $h$ is denoted by 

\begin{equation}\label{fork1}
 {p_s} = \frac{{{b_{h-1}}(m) + {v_{h-1}}(m)}}{w}.
\end{equation}
Considering that there are 
$\left\lfloor {\frac{{n + 1}}{2}} \right\rfloor  - 1$ no-turn nodes at Step $h$ to generate  no-turn blocks, the probability of a fork incurred by multiple no-turn nodes is thereby denoted by
\begin{equation}\label{fork2}
{p_f} = 1 - {(1 - {p_s})^{\left\lfloor {\frac{{n + 1}}{2}} \right\rfloor  - 1}}. 
\end{equation}

\noindent\textbf{\emph{Accurate delay range to prohibit forks}}: According to Eq.~(\ref{fork1}), a fork likely occurs when a no-turn node fails to consider the time required for the in-turn block broadcasting and validation (\emph{i.e.}, $b_{h} (m)+v_{h}(m)$). In other words, when the delay $x$, randomly selected by a no-turn node from the range of $(0,w)$  is less than the time cost for broadcasting and validation of the in-turn block, a fork occurs. 

%Therefore, a timer mechanism adopted by the no-turn nodes can mitigate forks effectively. 
To take this delay into account, we change the range of the random delay $x$ chosen by no-turn nodes to $(\beta,w)$. The fork probability after this change is %after adopting the timer mechanism is thereby changed from eq.(\ref{fork2}) to:
\begin{equation}
{p'_f} = 1 - {(1 - \frac{{{b_h}(m) + {v_h}(m)}-\beta}{w})^{\left\lfloor {\frac{{n + 1}}{2}} \right\rfloor  - 1}}.
\label{fork3}
\end{equation}
According to Eq.~(\ref{fork3}), the ideal case is ${p'_f}=0$ if we can set $b_h (m)+v_h (m)=\beta$.
%Thereby, each no-turn node should wait for at least $\beta$ to mitigate forks effectively. 
Each no-turn node can approximately measure $b_h(m)$ and $v_h(m)$ at the current Step $h$ based on the broadcasting and validation time cost of the last in-turn block generated from the same consensus node. Using the measured  $b_h(m)$ and $v_h(m)$, each no-turn node dynamically set $U(\beta, w)$ as the distribution to generate the random delay $x$.

\subsection{Minimizing $p_3$} \label{relationshiptpsapple}

Exceptional Case 3 may occur following the occurrence of a no-turn block from Exceptional Case 1. % is either followed by another no-turn block  (exceptional case 3)  or an in-turn block (exceptional case 2) }.
Due to the rigidly fixed order of in-turn nodes in Clique, there exists a ripple effect for occurrences of Exceptional Case 3. It means that a no-turn block committed at a step will trigger more no-turn blocks at the subsequent steps. Thus, Exceptional Case 3 occurs more frequently than Exceptional Cases 1 and 2. This conjecture is verified by the experiment presented in Fig.~\ref{process}(b), in which  $p_3$ grows faster than $p_1$ and $p_2$ with  $m$.

\begin{figure}[t]
  \centering
 \includegraphics[scale = 0.33]{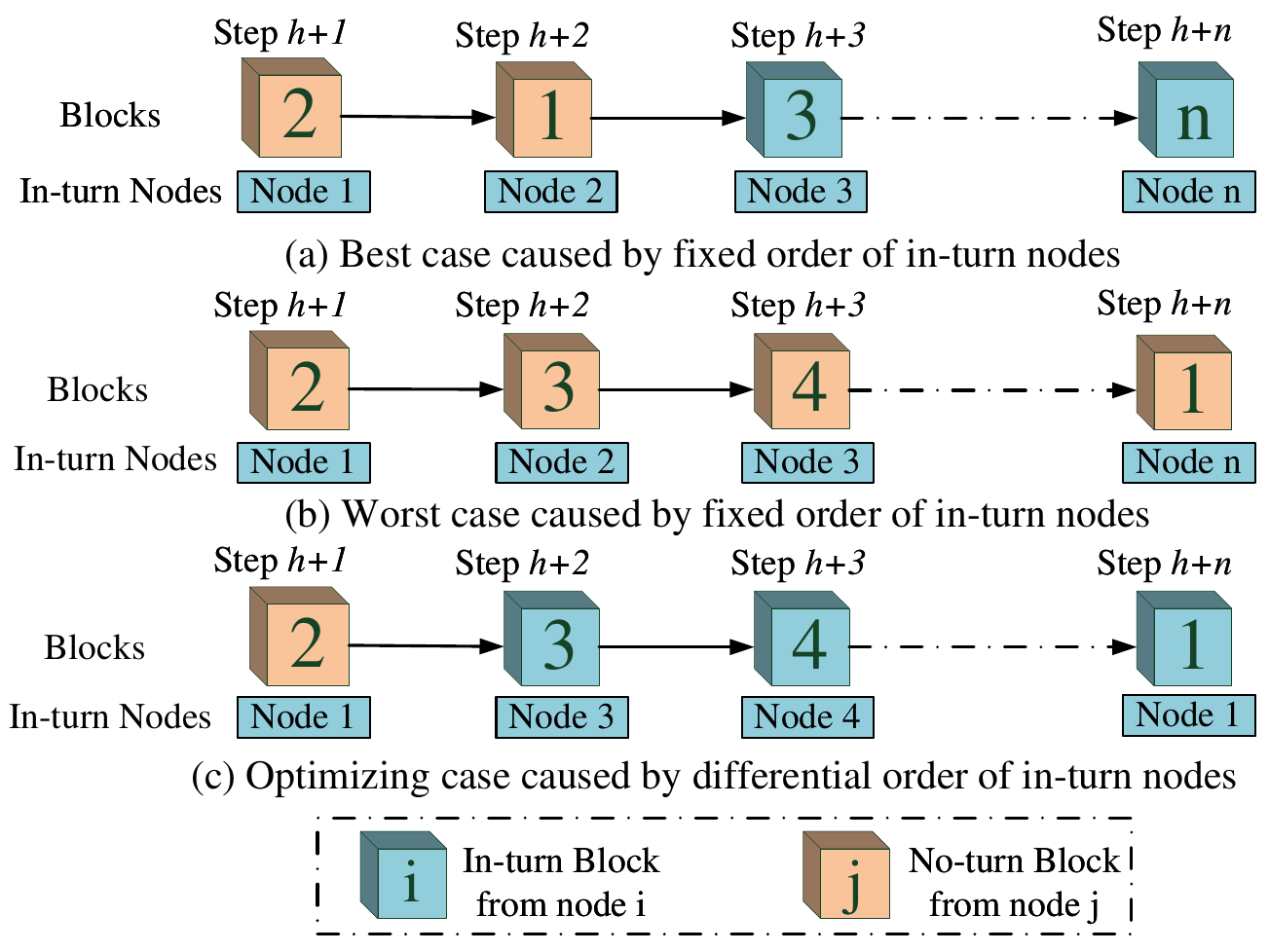}
  \caption{Different cases at multiple steps caused by different orders of in-turn nodes when a no-turn block is committed.}
%  \Description{A woman and a girl in white dresses sit in an open car.}
  \label{effect}
%\vspace{-10pt}
\end{figure}

%That is, one exceptional case 1 could trigger multiple exceptional case 3 so that as shown in  

 \noindent\textbf{\emph{Fixed order of in-turn nodes influencing $p_3$}}: %Typically, an exceptional case 3 is triggered when a no-turn block is selected after the no-turn block in the exceptional case 1. However, exceptional case 3 would be triggered frequently due to a "Ripple Effect" caused by the design of a fixed order of in-turn nodes in Clique. The “Ripple Effect” means that one no-turn block committed at a single step would trigger multiple no-turn blocks selected at the subsequent steps. 
To better understand the ripple effect in Clique due to the fixed/pre-determined order of in-turn nodes, we present two representative cases in Fig.~\ref{effect}(a) and Fig.~\ref{effect}(b), in which the order of in-turn nodes is $1, 2,\dots, n$. Fig.~\ref{effect}(a) presents the case without the ripple effect because the in-turn block is generated at Step $h+3$ following Exceptional Case 1 at Step $h+1$ and Exceptional Case 2 at Step $h+2$. In contrast, Fig.~\ref{effect}(b) shows the case with a significant ripple effect. In this case, a series of no-turn blocks are generated over multiple consecutive steps. 

By scrutinizing the difference between these two cases, we can see the reason for the ripple effect. Clique restricts that each node can only generate a single block over $\left\lfloor {\frac{n}{2}} \right\rfloor  + 1$  steps \cite{clique_EIP225,ethereum}. Node 3 has committed to generate a no-turn block at Step $h+2$, and thus it cannot generate the in-turn block at Step $h+3$, resulting in the generation of a no-turn block from node 4 at Step $h+3$.   However, a specific node that will commit to the generation of the no-turn block at Step $h+2$ cannot be centrally controlled, implying that the influence of the ripple effect can persist over multiple steps. 

%and worst case arising from the fixed order of in-turn blocks when an in-turn node fails at a step, as depicted in Fig. \ref{effect}(a) and Fig. \ref{effect}(b), respectively. The best case will only contain two no-turn blocks, while the worst case will contain infinite no-turn blocks. 

%For the best case as shown in Fig.\ref{effect}(a), when the in-turn node 1 at step $h+1$ fails, one of the no-turn nodes at this step will compete to generate a no-turn block committed to the global ledger. Since different no-turn nodes select and wait for different random delays, the winner is random. If the winner is node 2, it cannot generate the in-turn block at step $h+2$ due to the restriction that each node can only generate a block within 
%$\left\lfloor {\frac{n}{2}} \right\rfloor  + 1$ steps. Therefore, one of the no-turn nodes at step $h+2$ will compete to generate a no-turn block. If node 1 recovers from failure and becomes the winner to generate a no-turn block, the total number of no-turn blocks is 2. However, for the worst case as shown in Fig.\ref{effect}(b), if the winner generates no-turn block at each step is the in-turn node at the next step, there are infinite no-turn blocks committed to the global ledger. 

\noindent\textbf{\emph{Differential order of in-turn nodes}}: 
To prohibit the ripple effect in Clique, we propose a differential order of in-turn nodes.  In Clique, the in-turn node at Step $h$ is specified to the consensus node with index $i$ satisfying $h\%n=i$, where $n$ is the total number of consensus nodes. %Meanwhile, due to the restriction that each consensus node can only generate a block within $\left\lfloor {\frac{n}{2}} \right\rfloor  + 1$ steps to tolerate up to $\frac{n}{2}$ faulty nodes, Ripple Effect will be triggered if the in-turn node at the current step has generated a no-turn block at the past $\left\lfloor {\frac{n}{2}} \right\rfloor $ steps, as shown in Fig.\ref{effect}(a) and Fig.\ref{effect}(b). 
Rather than fixing indices of in-turn nodes,  ExClique dynamically sets the order of in-turn nodes as follows. Without loss of generality, we consider the scenario that in-turn node $i$ at Step $h+1$ fails to generate an in-turn block, and no-turn node $j$ is the winner to generate a no-turn block committed to the global ledger. The in-turn node at the subsequent  Step $h+2$ is then assigned to the consensus node with index $j+1$. 
%That is, 
 In other words, the index of the in-turn node at each step is purely determined by the index of the consensus node generating the block (regardless of no-turn or in-turn blocks) in the last step. 

Fig.~\ref{effect}(c) exemplifies the differential order of in-turn nodes in ExClique. If in-turn node 1 fails at Step $h+1$ and no-turn node 2 is the winner at this step,  node 3 becomes the in-turn node at Step $h+2$, and consensus node 4 becomes the in-turn node at Step $h+3$, and so on.

%Benefiting from the design of dynamic order, the Ripple Effect caused by the fixed order is effectively eliminated. However, this 
The differential order of ExClique introduces unfairness among consensus nodes because it can disrupt the income balance of different nodes. Specifically, when $j>i+1$, 
consensus nodes with indexes falling within the range $[i+1,j)$ will miss the opportunity to earn transaction fees (\emph{i.e.}, block rewards) by generating blocks during Steps $[h+i+1,h+i+n-j]$.
%if the in-turn nodes in these steps function normally. 
Here, we suppose that $j$ is the no-turn node generating the no-turn block after node $i$ generates a no-turn block in the last step. 

%\color{blue}
\noindent\textbf{\emph{Fair Smart Contract:}} To eliminate the unfairness caused by using the differential order in Clique, we propose a smart contract to distribute cumulative transaction
fees equally among multiple consensus nodes, \emph{e.g.}, a distributed contract used in BNB blockchain~\cite{BSC}. In other words, rather than sending transaction fees directly to block generators, transaction fees are temporarily accumulated within the fair smart contract. Then, the fees are distributed equally to active consensus nodes after a certain period.

%we have developed a fair smart contract inspired by \cite{BSC}, which can distribute cumulative transaction fees equally among multiple consensus nodes. In other words, rather than sending transaction fees directly to block generators, transaction fees are temporarily accumulated within the fair smart contract. Then, the fees are distributed equally to active consensus nodes after a certain period. 

Nonetheless, the smart contract for evenly distributing transaction fees should have a mechanism to prevent inactive or lazy nodes. Otherwise, inactive or lazy nodes can earn fees without generating any blocks. Besides, the smart contract takes a longer time to settle fees, which might be unacceptable for nodes who are eager to access funds promptly \cite{BSC}. Considering these issues, our design is based on a sliding window with length $n$ (the same as the number of nodes) and adopts the mechanism of uncle block in Ethereum \cite{uncle} to detect lazy nodes.  The fees are only distributed to active nodes periodically. The detailed design is presented in \emph{Algorithm \ref{alg1}}.
Specifically, a block is considered as a confirmed block if it is committed to the blockchain, and as an uncle block if it is not committed but contains correct information and is recorded in the confirmed block. Thus, a consensus node can be deemed active when it generates the uncle block even if its block is not committed. Then, for a given Step $h$,  consensus nodes not generating any confirmed/uncle blocks in the past $n$ steps\footnote{Each node has at least one chance to become an in-turn node within every $n$ steps based on the limit of block generation and the differential order.} are identified as lazy/inactive nodes. Lastly, the smart contract only transfers the rewards evenly to active nodes. 

\begin{algorithm}[t]
	\caption{Fair Smart Contract} 
	\label{alg1}
	
	\begin{algorithmic}[1]
						%\small  
\footnotesize
		\REQUIRE transaction fees $f$ for a confirmed block at Step $h$, number of consensus nodes $n$, two empty arrays $A_1$ and $A_2$
		\\    
		\ENSURE $A_2$	
		\STATE $i=0$
		%\STATE Remove the transactions in $G_c$ with timestamp later than $T$. 
		\WHILE{$i<n$} 
		\STATE  Clear $A_1$
            \STATE  Select the block generator at Step $h-n+i$ and add it in $A_1$
            \STATE  Detect all uncle blocks recorded in the confirmed block at Step $h-n+i$ and append their generators in $A_1$
            \STATE  Calculate the number of generators in $A_1$ and denote it as $a_1$
            \STATE  $j=0$
            \WHILE{ $j<a_1$}
		\IF{$A_1[j]$ is not in $A_2$} 
		\STATE Append $A_1[j]$ in $A_2$
		\ENDIF
            \STATE $j=j+1$
            \ENDWHILE
            \STATE $i=i+1$
            \ENDWHILE
            \STATE Calculate the number of generators in $A_2$ and denote it as $a_2$
            \STATE Transfer the transaction fees $f/a_2$ to each generator in $A_2$
	\end{algorithmic}
 %\vspace{-5pt}
\end{algorithm}

\color{black}
\section{Performance Evaluation}\label{experiment}
In this section, we implement  ExClique and deploy it in a permissioned blockchain network to evaluate its performance.
\subsection{Experimental Settings}\label{setup}
To verify and evaluate the performance of   ExClique, we implement the prototype of different Clique consensus algorithms by modifying Geth, which is the blockchain software used for implementing Clique \cite{ethereum}. 

\subsubsection{System Settings}
%\noindent\textbf{\emph{System Settings}}: 
For conducting our experiments, we build a  permissioned blockchain network environment using the docker container technology on a Linux server. In this environment, the server deploys multiple docker containers and allocates its hardware resources (\emph{e.g.}, CPU, RAM, and SSD) to containers. Each docker container is designated as a virtual ``physical machine" representing a location somewhere in the world, and network parameters between two containers are configured by geographical labels to simulate a distributed network. The specific blockchain software is loaded and executed within a docker container, serving as a consensus node. These consensus nodes then establish connections with each other to form a permissioned blockchain network. The specific parameters of the  permissioned blockchain network are shown in Table \ref{parameters}. 

\begin{table}[]
\centering
\caption{Parameters of  permissioned blockchain network}
  \label{parameters}
\begin{tabular}{|l|l|}
\hline
\textbf{Parameter}         & \textbf{Values}                         \\ \hline
Network Size $n$     & 21, 31, 41, 51, and 101      \\ \hline
Step Duration  $t_b$   & 3s                             \\ \hline
Topology          & Fully P2P Network              \\ \hline
Bandwidth         & 32 Mbit/s                      \\ \hline
Packet Loss Rate  & Random Value within (0, 0.1)   \\ \hline
Propagation Delay & Random Value within (0, 200ms) \\ \hline
\end{tabular}
\end{table}

\subsubsection{Baselines} 
%\noindent\textbf{\emph{Baselines}}: 
We compare the performance of ExClique with three baselines: 1)
Clique; 2) Modified Clique with BCB, \emph{i.e.}, Clique plus the compact block protocol that replaces full transactions with transaction hashes \cite{compact}; 3) Modified Clique with HCB, \emph{i.e.}, Clique plus hybrid compact block that replaces full transactions with short transaction hashes and embeds a prediction model into missing transactions to reduce additional communication delay \cite{hcb}.

\subsubsection{Evaluation Metrics}
%\noindent\textbf{\emph{Evaluation Metrics}}:
To evaluate the performance of  Exclique, we select the following metrics: 1) TPS performance to evaluate the speed to process transactions; 2) block size and block broadcasting time to verify the efficiency of PCB; 3) forks rate ($p_f$) and the rate of each exceptional case ($p_1$, $p_2$, and $p_3$) to verify the efficiency improvement by the accurate delay range and the differential order of in-turn nodes; and 4) block reward distribution on consensus nodes to verify the efficiency of the fair smart contract. %of accurate delay range and differential order of the in-turn nodes.

%{\bf YP: do not we need to talk about baslines and evaluation metrics here?}
\subsection{Experimental Results}
%We report our experimental results in this subsection. 
%This subsection begins by presenting experimental results aimed at verifying the efficiency of three schemes within the  ExClique framework. Specifically, we assess the block size and block broadcasting time for the TCB protocol, evaluate forks concerning the timer mechanism, and examine the no-turn block rate with respect to the dynamic order of in-turn nodes. Furthermore, we conduct an evaluation of the empty block rate and TPS performance of our  ExClique.

%\begin{figure}[h]
\begin{figure}[t]
  \centering
  \includegraphics[scale=0.47]{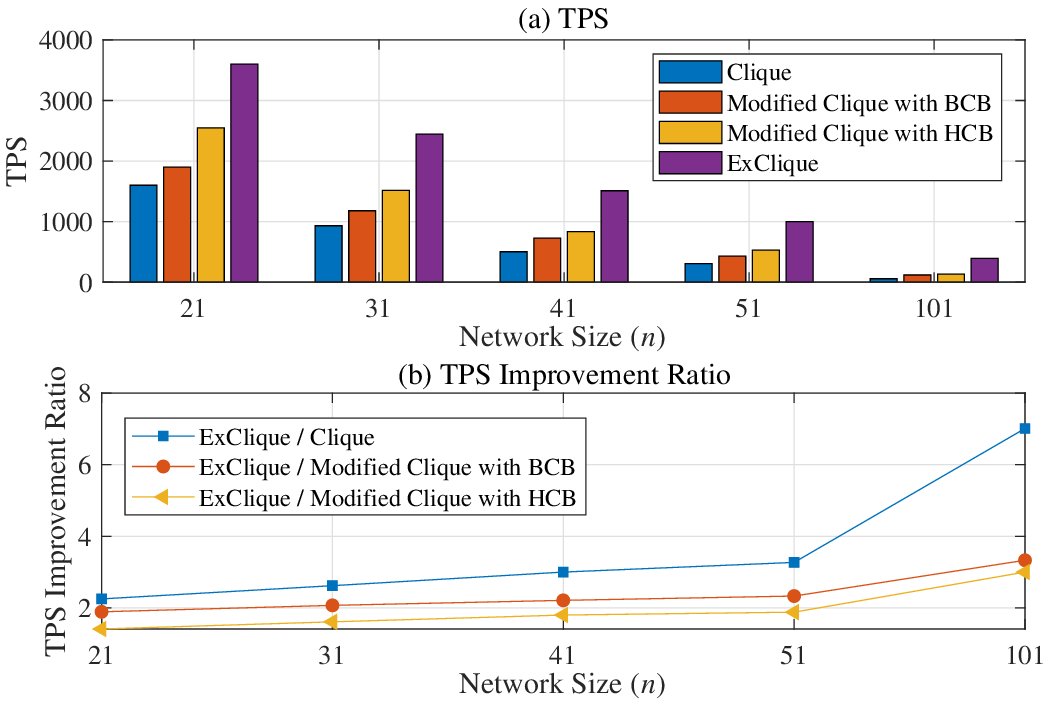}
  \caption{TPS performance of different Cliques under different network sizes.  }
%  \Description{A woman and a girl in white dresses sit in an open car.}
  \label{empty_block_rate_and_tps}
%\vspace{-10pt}
\end{figure}
\noindent\textbf{\emph{TPS Performance}}: %Three schemes in  ExClique are introduced to mitigate the factors restricting the TPS performance of Clique.
The network size, denoted as $n$, is a critical parameter significantly influencing TPS performance because it is difficult to reach a consensus when there are a large number of consensus nodes. For evaluating TPS, we vary the network size from 21 to 101 to study how much TPS can be improved by ExClique under different network scales. Given $n$, to find the optimal TPS for each consensus algorithm, we enumerate different values for $m$ to search $m^*$ satisfying Eq.~(\ref{idealTPS1}) to ensure that we compare the optimal TPS of different consensus algorithms. 
%We run an experiment with a specific network size and one of four consensus algorithms to perform 200 steps under a specific $m$ (number of transactions per block). By }  
%that determines key aspects of Clique and is a significant consideration for enterprises when deploying a permissioned blockchain, we conducted TPS performance measurements for four different Cliques across varying network sizes.

%On the other hand, because the network size $n$ is a parameter to determine the key parameters in Clique and the enterprises often need to consider the relationship between TPS and network size when deploying a permissioned blockchain, we measured the TPS performance of four Cliques under different network sizes. 

The experimental results are presented in Fig.~\ref{empty_block_rate_and_tps}, from which we can see that: 1) ExClique always achieves the highest TPS performance in comparison to other baselines, indicating its advantages in applications with massive transactions. 2) Compared with baselines, Clique achieves a higher TPS gain with a larger network scale. In particular, when the network size is 21 (which is the size adopted in \cite{heco,BSC}), Clique can improve TPS by 1.41$\times$-3.0$\times$ compared to baselines. However, when the network scale is increased to 101, the improvement is increased to 2.25$\times$-7.01$\times$.  
3)  Modified Clique with BCB and Modified Clique with HCB are two novel competitive baselines proposed in our work. They outperform Clique by compacting blocks. However, their performance is inferior to ours since ExClique designs the compact block protocol based on the characteristics of Clique-based blockchains and adopts additional strategies to prohibit exceptional cases.

\noindent\textbf{\emph{Block Size and Block Broadcasting Time}}: We conduct experiments to verify that ExClique can improve TPS because PCB in ExClique can shrink the block size effectively to reduce the block broadcasting time. In the experiment, we measure the average block size and block broadcasting time of each algorithm in the network with 21 consensus nodes. Here, the network size aligns with the configurations in real-world systems, \emph{e.g.}, \cite{heco, BSC}.
We vary $m$ from 0 to 12,350 per block to exhibit that PCB can consistently outperform baselines under various scenarios.

%\begin{figure}[h]
\begin{figure}[t]
  \centering
  \includegraphics[scale=0.35]{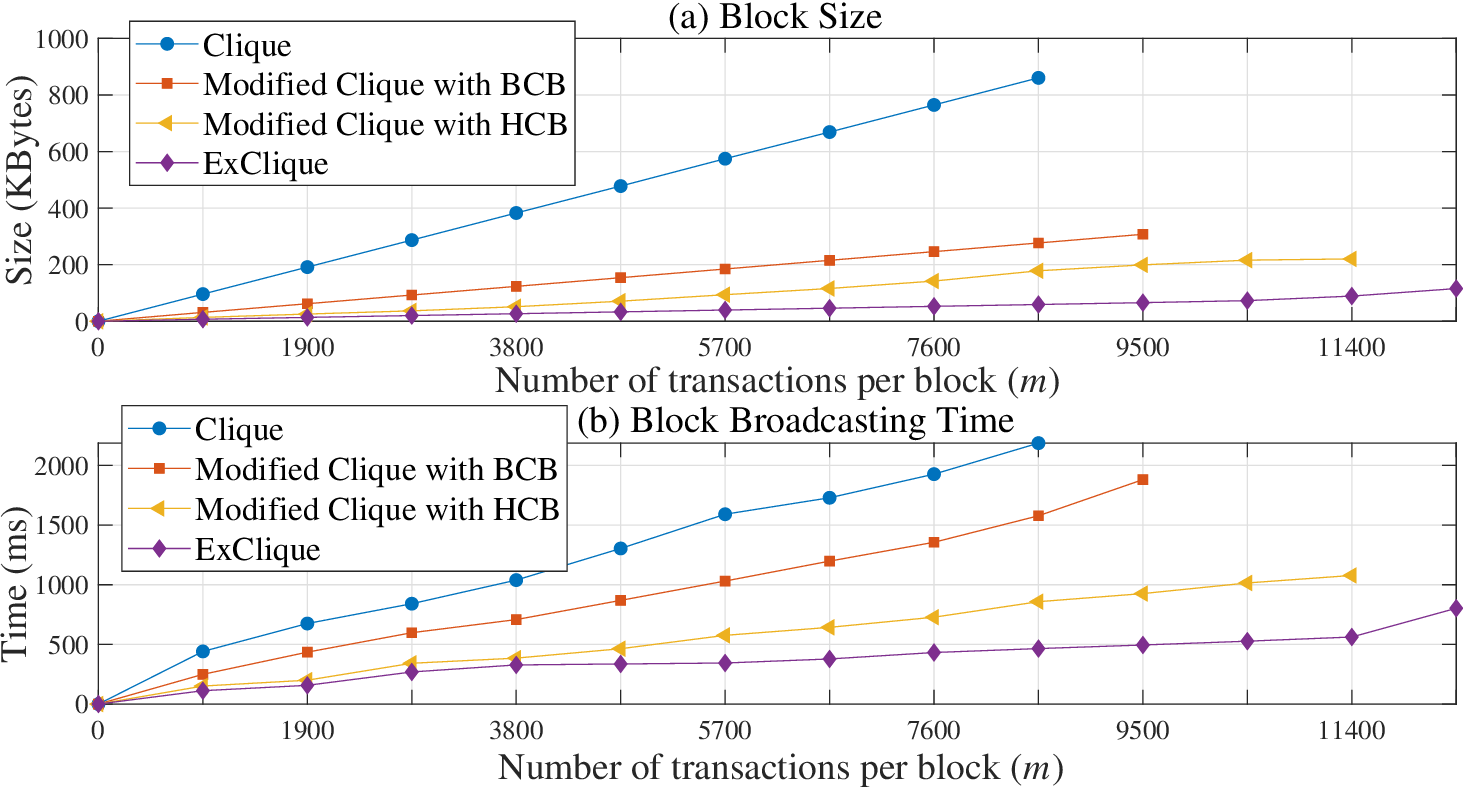}
  \caption{Block Size and Block Broadcasting Time for different Cliques in the network with 21 consensus nodes.}%: (a) Block Size; (b) Block Broadcasting Time.}
%  \Description{A woman and a girl in white dresses sit in an open car.}
  \label{block}
 % \vspace{-10pt}
\end{figure}

The experimental results are shown in Fig.~\ref{block}, where the horizontal axis represents the number of transactions per block (\emph{i.e.}, $m$).  %indicates the number of transactions per block (i.e., $m$). 
Fig.~\ref{block} exhibits the superb performance of the PCB protocol in ExClique. Notably, regardless of $m$, ExClique always achieves the smallest block size and shortest block broadcasting time among all four consensus algorithms. Besides, the improvement of ExClique becomes more pronounced as $m$ rises. When $m=8,550$, %(the maximum value that four Clique consensus algorithms work normally), 
the block size in ExClique is only around 1/14 that of Clique, 1/5 that of Modified Clique with BCB, and 1/3 that of Modified Clique with HCB. Similarly, the block broadcasting time of ExClique is approximately 1/5 that of Clique, 1/3 that of Modified Clique with BCB, and 5/9 that of Modified Clique with HCB. 
In Fig.~\ref{block}, it is worth noting that the block size can significantly restrict the transaction process capacity of the blockchain. When $m=12,350$, ExClique can still work normally. In contrast, the system using Clique fails to work when $m>8,550$.  Modified Clique with BCB and Modified Clique with HCB fail to work when $m>9,500$ and $m>11,400$, respectively. These results underscore the remarkable efficiency gains achieved by our PCB in ExClique.

\noindent\textbf{\emph{Rate of Different Exceptional Cases and Fork}}: Our ExClique employs an accurate delay range to reduce fork rates $p_f$ to alleviate the occurrences of Exceptional Cases 1 and 2 (\emph{i.e.,} decrease $p_1$ and $p_2$), and uses the differential order of in-turn nodes to avoid Exceptional Case 3 (\emph{i.e.,} $p_3=0$). To illustrate the effectiveness of these two mechanisms, we measured $p_f$, $p_1$, $p_2$, and $p_3$ in the typical blockchain network with 21 consensus nodes by varying different $m$. 

\begin{figure}[t]
  \centering
\includegraphics[scale=0.45]{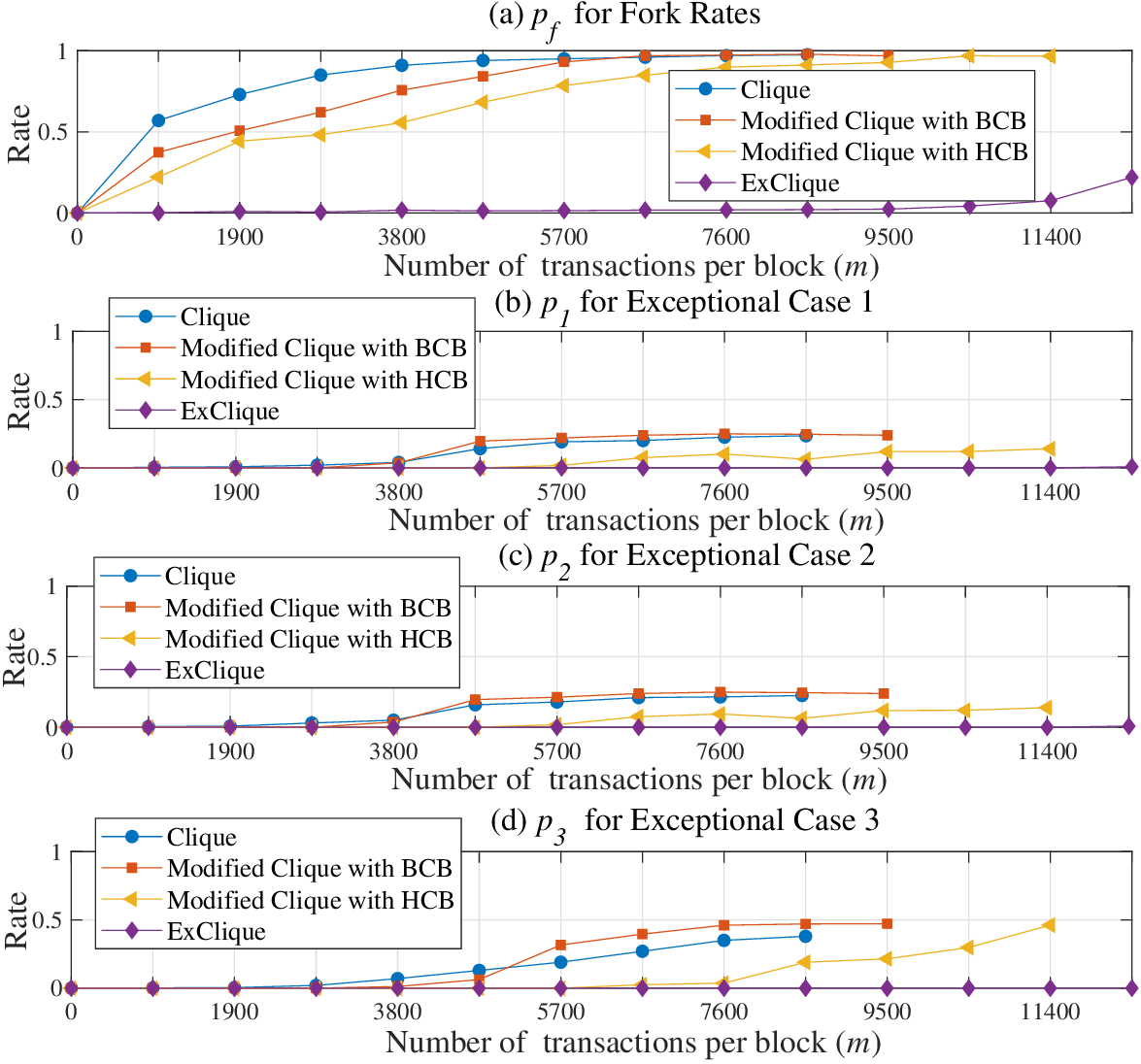}
  \caption{Rate of different exceptional cases and forks for different Cliques in the network with 21 consensus nodes. }
%  \Description{A woman and a girl in white dresses sit in an open car.}
  \label{forks_rate_and_no_turn_block_rate}
 % \vspace{-10pt}
\end{figure}

The experimental results are presented in Fig.~\ref{forks_rate_and_no_turn_block_rate}, with the horizontal axis representing the number of transactions per block ($m$). From Fig.~\ref{forks_rate_and_no_turn_block_rate}(a), the result reveals a significant difference in $p_f$ between our ExClique and other baselines. For baselines, their rates $p_f$ surge rapidly with respect to $m$, and all exceed 0.5 when $m$ reaches 5,700. Due to their high $p_f$, it is not difficult to understand that exceptional cases will occur more frequently with higher $p_1$ and $p_2$, which has been shown in Fig.~\ref{forks_rate_and_no_turn_block_rate}(b) and Fig.~\ref{forks_rate_and_no_turn_block_rate}(c).  Conversely, in  ExClique, the fork rates remain under 0.1, even when $m$ is set to 11,400, which can greatly reduce the occurrence odds of Exceptional Cases 1 and 2. %Thanks to the low fork rate of ExClique, both $p_1$ and $p_2$ can remain close to 0. 
Meanwhile, as shown in Fig. \ref{forks_rate_and_no_turn_block_rate}(d), with the help of the differential order of in-turn nodes, the ripple effect causing consecutive occurrences of Exceptional Case 3 is completely averted, making $p_3$ always equal to 0. Therefore,  our  ExClique can considerably enhance TPS by effectively prohibiting occurrences of exceptional cases.% in Section \ref{relationshiptpsapple}.

Besides, it is worth noting that prior investigations \cite{security_forks_pow,bbp} have consistently reported fork rates less than 0.1 in prominent blockchain networks such as Bitcoin, Ethereum 1.0, Litecoin, and Dogecoin. This steadfast low fork rate has been instrumental in upholding these networks' robust security and unwavering stability. Thus, our  ExClique offers a robust solution that maintains security even as $m$ is extremely large, \emph{e.g.,}  11,400, ultimately improving TPS. In contrast, baselines struggle to achieve high system security when $m>950$, highlighting the perspective of  ExClique in practice. %$ensuring both scalability and security.

\begin{figure}[t]
  \centering
  \includegraphics[width=\linewidth]{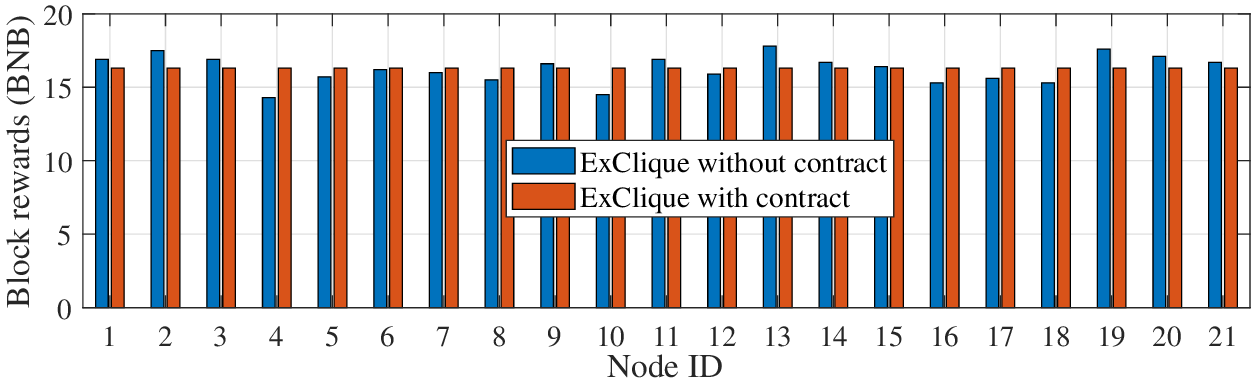}
  \caption{ Block rewards earned by each node in the network with 21 consensus nodes when ExClique adopts or not adopt the fair smart contract.  }
%  \Description{A woman and a girl in white dresses sit in an open car.}
  \label{block_rewards}
%  \vspace{-10pt}
\end{figure}

%\color{blue}
\noindent\textbf{\emph{Block Reward Distribution on Consensus Nodes}}: As discussed in Section \ref{relationshiptpsapple}, the differential order of ExClique can cause unfairness among consensus nodes, disrupting the income balance across different nodes. To address this issue, we design a fair smart contract. To validate the effectiveness of our method, we measure block rewards for each consensus node in a typical network with 21 consensus nodes, comparing ExClique with/without the smart contract. %versus when it does not. 
In each experiment case, 2,100 blocks are generated, mirroring the block reward distribution of the BNB blockchain \cite{BSC}.

The experimental results are shown in Fig.~\ref{block_rewards}, with the x-axis representing node ID. We can observe that ExClique, when adopting the fair smart contract, maintains a more balanced income distribution among different nodes compared to the case without the smart contract. %when the contract is not adopted. 
The results shed light on that  implementing the fair smart contract can effectively mitigate unfairness caused by the differential order. In contract, if the fair smart contract is absent, nodes  had significant variations in block rewards. %. without the contract now show reduced variance. 
 This validates that the fair smart contract effectively addresses the income imbalance and enhances the fairness of block reward distribution in ExClique. % consensus mechanism.

\color{black}
\section{Conclusion}\label{conclusion}
Clique stands as a prominent implementation of PoA. %, a significant alternative beyond the original PoW consensus algorithm in Ethereum. 
It was designed with a focus on energy efficiency, rendering it highly suitable for deployment in blockchain networks. 
However, the low transaction processing speed in large Clique has been overlooked in existing works, which has significantly restricted the use of Clique with massive transactions. 
Our work amends this deficiency by proposing ExClique, which improves TPS performance by designing: 1) a proactive compact block protocol to minimize the delay of block broadcasting based on the predictable in-turn node at each step; 2) a tighter delay range to diminish the fork rate; and 3) a differential order of in-turn nodes to prohibit no-turn block occurrences. Importantly, ExClique is friendly for implementation and compatible with the current blockchain systems adopting Clique. % without causing heavy additional burden.
%a significant challenge arises when multiple consensus nodes are allowed to generate blocks concurrently, leading to forks that can compromise system security and TPS performance. Additionally, the "Ripple Effect" and prolonged block broadcasting times limit TPS capabilities. To address these issues, we introduce ExClique, which incorporates three critical strategies: a Timer Mechanism to reduce forks, a Dynamic Order of In-Turn Nodes to mitigate the Demoni Effect, and a TCB Protocol to speed up block broadcasting. Comparative analysis with Clique reveals 
At last, extensive experiments are conducted to verify the superiority of ExClique, which reduces the fork rate to almost 0. Moreover, it substantially improves TPS by 2.25$\times$ in a blockchain network with 21 consensus nodes and an impressive 7.01$\times$ in a larger network with 101 consensus nodes. The superb performance of ExClique can advance blockchains by widening their applications in large-scale networks. %in terms of forks and TPS not only bolsters the robustness of blockchain systems but also enhances their ability to support larger-scale applications.

%{\bf YP: what is the future work? Can we outlook and briefly discuss what is the future work reseachers can do following our work? }

%A link inference model is proposed to infer a link between two nodes in Ethereum network by setting up a measurement node to perform a 5-step inference task. Based on the link inference model, a distributed measurement architecture consisting of multiple measurement nodes is further designed to connect more Ethereum nodes to discover a complete network topology. Compared to the existing schemes, DEthna achieves both high precision and recall. DEthna discover the topology of Goerli in two periods, which show more than half of NAT nodes in the network would lead to the “long tail” latencies to cover the entire network and a weak robustness to defend the targeted attack. 
%work %\section*{Acknowledgment}

%The preferred spelling of the word ``acknowledgment'' in America is without 
%an ``e'' after the ``g''. Avoid the stilted expression ``one of us (R. B. 
%G.) thanks $\ldots$''. Instead, try ``R. B. G. thanks$\ldots$''. Put sponsor 
%acknowledgments in the unnumbered footnote on the first page.

\clearpage

\vspace{12pt}

\bibliographystyle{IEEEtran}
%\bibliography{references}{}
%\bibliographystyle{splncs04} %这是一个样式文件
\bibliography{DEthna} %其中reference为reference.bib文件。

\begin{thebibliography}{10}
\providecommand{\url}[1]{#1}
\csname url@samestyle\endcsname
\providecommand{\newblock}{\relax}
\providecommand{\bibinfo}[2]{#2}
\providecommand{\BIBentrySTDinterwordspacing}{\spaceskip=0pt\relax}
\providecommand{\BIBentryALTinterwordstretchfactor}{4}
\providecommand{\BIBentryALTinterwordspacing}{\spaceskip=\fontdimen2\font plus
\BIBentryALTinterwordstretchfactor\fontdimen3\font minus
  \fontdimen4\font\relax}
\providecommand{\BIBforeignlanguage}[2]{{%
\expandafter\ifx\csname l@#1\endcsname\relax
\typeout{** WARNING: IEEEtran.bst: No hyphenation pattern has been}%
\typeout{** loaded for the language `#1'. Using the pattern for}%
\typeout{** the default language instead.}%
\else
\language=\csname l@#1\endcsname
\fi
#2}}
\providecommand{\BIBdecl}{\relax}
\BIBdecl

\bibitem{bitcoin}
S.~Nakamoto, ``Bitcoin: A peer-to-peer electronic cash system,''
  \emph{Decentralized Business Review}, 2008.

\bibitem{ethereum}
(2015) Go-ethereum. \url{https://github.com/ethereum/go-ethereum}.

\bibitem{web3}
L.~V. Kiong, \emph{Web3 Made Easy: A Comprehensive Guide to Web3: Everything
  you need to know about Web3, Blockchain, DeFi, Metaverse, NFT and
  GameFi}.\hskip 1em plus 0.5em minus 0.4em\relax Liew Voon Kiong, 2022.

\bibitem{web3_account}
T.~Wang, S.~Zhang, Q.~Yang, and S.~C. Liew, ``Account service network: A
  unified decentralized web 3.0 portal with credible anonymity,'' \emph{IEEE
  Network}, vol.~37, no.~6, pp. 101--108, 2023.

\bibitem{mev}
K.~Qin, L.~Zhou, and A.~Gervais, ``Quantifying blockchain extractable value:
  How dark is the forest?'' in \emph{2022 IEEE Symposium on Security and
  Privacy (SP)}.\hskip 1em plus 0.5em minus 0.4em\relax IEEE, 2022, pp.
  198--214.

\bibitem{nft}
L.~Ante, ``Non-fungible token (nft) markets on the ethereum blockchain:
  Temporal development, cointegration and interrelations,'' \emph{Economics of
  Innovation and New Technology}, pp. 1--19, 2022.

\bibitem{IoT}
F.~Chen, J.~Wang, C.~Jiang, T.~Xiang, and Y.~Yang, ``Blockchain based
  non-repudiable iot data trading: Simpler, faster, and cheaper,'' in
  \emph{IEEE INFOCOM 2022 - IEEE Conference on Computer Communications}, 2022,
  pp. 1958--1967.

\bibitem{duan2021metaverse}
H.~Duan, J.~Li, S.~Fan, Z.~Lin, X.~Wu, and W.~Cai, ``Metaverse for social good:
  A university campus prototype,'' in \emph{Proceedings of the 29th ACM
  International Conference on Multimedia}, 2021, pp. 153--161.

\bibitem{weak_consensus}
Q.~Wang and R.~Li, ``A weak consensus algorithm and its application to
  high-performance blockchain,'' in \emph{IEEE INFOCOM 2021 - IEEE Conference
  on Computer Communications}, 2021, pp. 1--10.

\bibitem{pow_age}
L.~Shi, T.~Wang, J.~Li, S.~Zhang, and S.~Guo, ``Pooling is not favorable:
  Decentralize mining power of pow blockchain using age-of-work,'' \emph{IEEE
  Transactions on Cloud Computing}, vol.~11, no.~3, pp. 2756--2769, 2022.

\bibitem{consensus_survey}
\BIBentryALTinterwordspacing
J.~Xu, C.~Wang, and X.~Jia, ``A survey of blockchain consensus protocols,''
  \emph{ACM Comput. Surv.}, vol.~55, no. 13s, jul 2023. [Online]. Available:
  \url{https://doi.org/10.1145/3579845}
\BIBentrySTDinterwordspacing

\bibitem{consensus_1}
\BIBentryALTinterwordspacing
D.~P. Oyinloye, J.~S. Teh, N.~Jamil, and M.~Alawida, ``Blockchain consensus: An
  overview of alternative protocols,'' \emph{Symmetry}, vol.~13, no.~8, 2021.
  [Online]. Available: \url{https://www.mdpi.com/2073-8994/13/8/1363}
\BIBentrySTDinterwordspacing

\bibitem{comparative}
M.~M. Islam, M.~M. Merlec, and H.~P. In, ``A comparative analysis of
  proof-of-authority consensus algorithms: Aura vs clique,'' in \emph{2022 IEEE
  International Conference on Services Computing (SCC)}.\hskip 1em plus 0.5em
  minus 0.4em\relax IEEE, 2022, pp. 327--332.

\bibitem{clique_EIP225}
P.~Szil{\'a}gyi. (2017) Eip-225: Clique proof-of-authority consensus protocol.
  \url{https://eips.ethereum.org/EIPS/eip-225}.

\bibitem{heco}
(2020) Heco chain. \url{https://github.com/stars-labs/heco-chain}.

\bibitem{BSC}
Binance. (2024) Binance smart chain.
  \url{https://docs.bnbchain.org/docs/learn/intro/}.

\bibitem{amazon}
A.~Flores. (2018) Launch enterprise-ready blockchain networks on aws in minutes
  with kaleido—a consensys solution.
  \url{https://aws.amazon.com/cn/blogs/apn/}.

\bibitem{clones_attack}
E.~Parinya, G.~Vincent, and J.~Guillaume, ``The attack of the clones against
  proof-of-authority,'' in \emph{Network and Distributed Systems Security
  (NDSS) Symposium}, 2020, pp. 1--14.

\bibitem{unfairness_exploring}
Q.~Wang, R.~Li, Q.~Wang, S.~Chen, and Y.~Xiang, ``Exploring unfairness on proof
  of authority: Order manipulation attacks and remedies,'' in \emph{Proceedings
  of the 2022 ACM on Asia Conference on Computer and Communications Security},
  2022, pp. 123--137.

\bibitem{Clique_Adv}
\BIBentryALTinterwordspacing
c.~wackerow, minimalism and 0x0OZ, ``Proof-of-authority (poa).'' [Online].
  Available:
  \url{https://ethereum.org/zh/developers/docs/consensus-mechanisms/poa/}
\BIBentrySTDinterwordspacing

\bibitem{fabric}
S.~Gupta, J.~Hellings, S.~Rahnama, and M.~Sadoghi, ``Building high throughput
  permissioned blockchain fabrics: challenges and opportunities,''
  \emph{Proceedings of the VLDB Endowment}, vol.~13, no.~12, 2020.

\bibitem{privateBlockchain}
T.~T.~A. Dinh, J.~Wang, G.~Chen, R.~Liu, B.~C. Ooi, and K.-L. Tan,
  ``Blockbench: A framework for analyzing private blockchains,'' in
  \emph{Proceedings of the 2017 ACM International Conference on Management of
  Data}, 2017, pp. 1085--1100.

\bibitem{chaos_engineer}
S.~Sondhi, S.~Saad, K.~Shi, M.~Mamun, and I.~Traore, ``Chaos engineering for
  understanding consensus algorithms performance in permissioned blockchains,''
  in \emph{2021 IEEE Intl Conf on Dependable, Autonomic and Secure Computing,
  Intl Conf on Pervasive Intelligence and Computing, Intl Conf on Cloud and Big
  Data Computing, Intl Conf on Cyber Science and Technology Congress
  (DASC/PiCom/CBDCom/CyberSciTech)}.\hskip 1em plus 0.5em minus 0.4em\relax
  IEEE, 2021, pp. 51--59.

\bibitem{performance_evaluation_blockchain}
\BIBentryALTinterwordspacing
A.~Ahmad, A.~Alabduljabbar, M.~Saad, D.~Nyang, J.~Kim, and D.~Mohaisen,
  ``Empirically comparing the performance of blockchain's consensus
  algorithms,'' \emph{IET Blockchain}, vol.~1, no.~1, pp. 56--64, 2021.
  [Online]. Available:
  \url{https://ietresearch.onlinelibrary.wiley.com/doi/abs/10.1049/blc2.12007}
\BIBentrySTDinterwordspacing

\bibitem{oneline_game}
L.~Xiao, ``Blockchain in online games and what can be learned from it,''
  \emph{Frontiers in Business, Economics and Management}, vol.~12, no.~3, pp.
  35--38, 2023.

\bibitem{permission3}
D.~Agrawal, A.~El~Abbadi, M.~J. Amiri, S.~Maiyya, and V.~Zakhary, ``Blockchains
  and databases: Opportunities and challenges for the permissioned and the
  permissionless,'' in \emph{Advances in Databases and Information Systems:
  24th European Conference, ADBIS 2020, Lyon, France, August 25--27, 2020,
  Proceedings 24}.\hskip 1em plus 0.5em minus 0.4em\relax Springer, 2020, pp.
  3--7.

\bibitem{CAP_analysis_1}
S.~D. Angelis, L.~Aniello, R.~Baldoni, F.~Lombardi, A.~Margheri, and
  V.~Sassone, ``{PBFT} vs proof-of-authority: Applying the {CAP} theorem to
  permissioned blockchain,'' in \emph{{ITASEC}}, ser. {CEUR} Workshop
  Proceedings, vol. 2058.\hskip 1em plus 0.5em minus 0.4em\relax CEUR-WS.org,
  2018.

\bibitem{choice_ethereum_client}
C.~N. Samuel, S.~Glock, F.~Verdier, and P.~Guitton-Ouhamou, ``Choice of
  ethereum clients for private blockchain: Assessment from proof of authority
  perspective,'' in \emph{2021 IEEE International Conference on Blockchain and
  Cryptocurrency (ICBC)}.\hskip 1em plus 0.5em minus 0.4em\relax IEEE, 2021,
  pp. 1--5.

\bibitem{deadlock}
M.~Martínez. (2019) Poa network, all the sealers are waiting for each other
  after 2 months running, possible deadlock?
  \url{https://github.com/ethereum/go-ethereum/issues/18402}.

\bibitem{idchain}
A.~Symons. (2021) Deadlock resolver.
  \url{https://github.com/IDChain-eth/IDChain/blob/release/1.9/deadlock_resolver.py}.

\bibitem{ethereumYellow}
\BIBentryALTinterwordspacing
G.~Wood \emph{et~al.}, ``Ethereum: A secure decentralised generalised
  transaction ledger,'' \emph{Ethereum Project Yellow Paper}, vol. 151, no.
  2014, pp. 1--32, 2014. [Online]. Available:
  \url{https://cryptodeep.ru/doc/paper.pdf}
\BIBentrySTDinterwordspacing

\bibitem{hcb}
C.~Zhao, T.~Wang, S.~Zhang, and S.~C. Liew, ``Hcb: Enabling compact block in
  ethereum network with secondary pool and transaction prediction,'' \emph{IEEE
  Transactions on Network Science and Engineering}, vol.~11, no.~1, pp.
  1077--1092, 2024.

\bibitem{cbfCaclulate}
L.~Fan, P.~Cao, J.~Almeida, and A.~Z. Broder, ``Summary cache: a scalable
  wide-area web cache sharing protocol,'' \emph{IEEE/ACM Transactions on
  Networking}, vol.~8, no.~3, pp. 281--293, 2000.

\bibitem{similarity}
K.~Dae-Yong, E.~Meryam, and J.~Hongtaek, ``Examining bitcoin mempools
  resemblance using jaccard similarity index,'' in \emph{2020 21st Asia-Pacific
  Network Operations and Management Symposium (APNOMS)}.\hskip 1em plus 0.5em
  minus 0.4em\relax IEEE, 2020, pp. 287--290.

\bibitem{compact_1}
Z.~Hu and Z.~Xiao, ``Dino: A block transmission protocol with low bandwidth
  consumption and propagation latency,'' in \emph{IEEE INFOCOM 2022 - IEEE
  Conference on Computer Communications}, 2022, pp. 1319--1328.

\bibitem{compact}
M.~Corallo. (2016) Compact block relay.
  \url{https://github.com/bitcoin/bips/blob/master/bip-0152.mediawiki}.

\bibitem{bbp}
\BIBentryALTinterwordspacing
C.~Zhao, S.~Zhang, T.~Wang, and S.~C. Liew, ``Bodyless block propagation: Tps
  fully scalable blockchain with pre-validation,'' \emph{Future Generation
  Computer Systems}, vol. 163, p. 107516, 2025. [Online]. Available:
  \url{https://www.sciencedirect.com/science/article/pii/S0167739X24004801}
\BIBentrySTDinterwordspacing

\bibitem{bloom_survey}
A.~Broder and M.~Mitzenmacher, ``Network applications of bloom filters: A
  survey,'' \emph{Internet Mathematics}, vol.~1, no.~4, pp. 485--509, 2004.

\bibitem{false_negative_problem}
D.~Guo, Y.~Liu, X.~Li, and P.~Yang, ``False negative problem of counting bloom
  filter,'' \emph{IEEE Transactions on Knowledge and Data Engineering},
  vol.~22, no.~5, pp. 651--664, 2010.

\bibitem{transaction_size}
J.~K. lumierre, Linmao~Song. (2018) Transaction size.
  \url{https://ethereum.stackexchange.com/questions/30175/}.

\bibitem{uncle}
\BIBentryALTinterwordspacing
Y.~Huang, B.~Wang, and Y.~Wang, ``Research and application of smart contract
  based on ethereum blockchain,'' \emph{Journal of Physics: Conference Series},
  vol. 1748, no.~4, p. 042016, jan 2021. [Online]. Available:
  \url{https://dx.doi.org/10.1088/1742-6596/1748/4/042016}
\BIBentrySTDinterwordspacing

\bibitem{security_forks_pow}
A.~Gervais, G.~O. Karame, K.~W{\"u}st, V.~Glykantzis, H.~Ritzdorf, and
  S.~Capkun, ``On the security and performance of proof of work blockchains,''
  in \emph{Proceedings of the 2016 ACM SIGSAC Conference on Computer and
  Communications Security}, 2016, pp. 3--16.

\end{thebibliography}
\end{document}